%% file: main.tex
\documentclass[ALICE,manyauthors]{cernphprep}
\usepackage[comma,square,numbers,sort&compress]{natbib}
\usepackage{hyperref}
\usepackage{lineno}
\usepackage{xspace}
\usepackage{xcolor}
\usepackage[T1]{fontenc}
\usepackage{orcidlink} 

\begin{document}
\input{commands.tex}

\begin{titlepage}
\PHyear{2026}       
\PHnumber{062}      
\PHdate{10 March}  

  \title{Production of \Xis and \Oms hyperons in high-multiplicity \\ proton-proton collisions at $\mathbf{\sqrt{\textit{s}}}$ = 13~TeV}
  \ShortTitle{$\Xi$ and $\Omega$ production in high-multiplicity pp at 13 TeV}    

\Collaboration{ALICE Collaboration\thanks{See Appendix~\ref{app:collab} for the list of collaboration members}}
\ShortAuthor{ALICE Collaboration} 

\begin{abstract}
This paper presents the first measurements of \Xis and \Oms hyperon yields at the highest multiplicities reached in pp collisions at $\sqrt{s}=13$~TeV.
This measurement exploits the high-multiplicity pp collisions collected by ALICE with dedicated triggers. The selected collisions are characterised by about 30 charged particles per unit of rapidity, over four times more than in minimum-bias pp collisions at the same centre-of-mass energy, and about twice as many as in minimum-bias p--Pb or very peripheral Pb--Pb collisions at similar energies.
The production yields and average transverse momenta of the hyperons agree with trends indicated by previous measurements in pp collisions at lower multiplicities. The difference in average transverse momenta between pp and p--Pb collisions, observed with the new high-multiplicity pp data, provides additional insight into the underlying particle production mechanisms in small systems. The results support a strong correlation between multi-strange hadron production and final-state multiplicity regardless of the collision system at the LHC energies, extending this observation to the highest multiplicity reached in pp collisions.
The comparison with several state-of-the-art models (\textsc{Pythia}8.2 with the Monash 2013 tune, \textsc{Pythia}8.2 with Ropes, and EPOS4) suggests that the description of strange-hadron production is improved by recently introduced features such as interactions between overlapping strings in \textsc{Pythia}8.2 with Ropes and the collective expansion of high-density string regions in EPOS4.

\end{abstract}
\end{titlepage}

\setcounter{page}{2} 


\section{Introduction}
Historically, one of the first suggested signatures of quark--gluon plasma (QGP) formation in heavy-ion collisions was the enhanced production of strange hadrons with respect to minimum-bias (MB) proton-proton (pp) collisions~\cite{PhysRevLett.48.1066, KOCH1983151, KOCH1986167}.
This strangeness enhancement was first measured at the SPS in various nucleus--nucleus collision systems~\cite{Bartke1990, WA97, Antinori:2006ij, Alt:2004kq, PhysRevC.78.034918, PhysRevC.80.034906}, then in Au--Au collisions at RHIC~\cite{PhysRevC.77.044908} and in Pb--Pb collisions at the LHC~\cite{PbPb5TeV, PbPb276TeV}.
A relative enhancement of strange to non-strange hadron yields was also observed in high-multiplicity pp collisions compared to minimum-bias ones by the ALICE experiment at the LHC~\cite{2017, 7TeV, ALICE:2019avo, pp5TeV, ALICE:2025aqz, ALICE:2024zxp}. 
The results show that the strange to non-strange hadron yield ratios increase as a function of the final-state charged-particle multiplicity smoothly across collision systems, from low multiplicity pp collisions up to high multiplicity central Pb--Pb collisions, where saturation occurs~\cite{pPb5TeV, PbPb5TeV}. This increase with multiplicity is found to be more pronounced for hadrons with a larger strangeness content, making multi-strange hyperons a particularly sensitive probe.
Moreover, no dependence on the collision energy is observed at the LHC energies~\cite{ALICE:2022wpn}.
These findings suggest a common particle production mechanism across different collision systems, and indicate a strong correlation between strange-to-non-strange hadron production and final-state particle multiplicity, regardless of the collision system.
To further investigate the correlation between strangeness enhancement and final-state multiplicity, the measurement presented in this paper addresses strange hadron production in pp collisions characterised by multiplicity values twice as large as those in the most peripheral analysed Pb–Pb collisions. This allows a comparison of strangeness production in systems of very different sizes at similar multiplicities.

\sloppy This article presents the measurement of multi-strange hadron yields ($\xim, \xip, \omm, \omp$) in high-multiplicity pp collisions at $\sqrt{s} =$~13~TeV collected by the ALICE detector during the LHC Run~2 from 2016 to 2018.
By making use of a trigger dedicated to the selection of high-multiplicity pp events, the hyperon yields are measured for the first time in pp collisions characterised by at least 30 charged particles produced per unit of rapidity at midrapidity, up to five times the value measured in a minimum-bias pp collision, and comparable to the one reached in high-multiplicity p--Pb collisions and in peripheral Pb--Pb collisions.
The hyperon yields are studied as a function of their transverse momentum and the event charged-particle multiplicity.
Strange to non-strange hadron ratios are compared to previous results in small and large collision systems.
In addition, the results are compared to state of the art QCD-inspired models, namely \textsc{Pythia}8.2 with the Monash 2013 tune~\cite{Sjostrand:2014zea}, \textsc{Pythia}8.2 with Ropes~\cite{Bierlich:2014xba}, and EPOS4~\cite{EPOS4_PhysRevC.108.034904, EPOS4_PhysRevC.108.064903, EPOS4_PhysRevC.109.014910}, as well as with the canonical statistical model~\cite{Vovchenko:2019kes, Vovchenko:2019pjl}.

The paper is organised as follows.
In Sec.~\ref{sec:exp}, the experimental setup and the data sample used for this measurement are presented; in Sec.~\ref{sec:analysis}, the experimental details of this study are outlined along with the associated systematic uncertainties.
Sec.~\ref{sec:results} presents the \pt spectra, the \pt-integrated yields, and the \meanpt of \Xis and \Oms as a function of the charged-particle multiplicity, and their comparison with different model predictions.
Finally, the conclusions are drawn in Sec.~\ref{sec:conclusions}.

\section{Experimental setup and data selection}
\label{sec:exp}
The ALICE detector~\cite{ALICEexp, ALICE_performance} is designed to perform tracking down to low momentum in the high-occupancy environment of heavy-ion collisions and provide excellent particle identification.
At midrapidity, in a solenoidal magnetic field of 0.5~T, the full azimuth is covered by the Inner Tracking System (ITS)~\cite{ITS} and the Time Projection Chamber (TPC)~\cite{TPC}, which are the primary tracking and particle identification detectors.
These detectors provide full track reconstruction in the pseudorapidity region $\lvert \eta \rvert < 0.9$ and offer reliable particle identification up to $p_\mathrm{T}\sim 20$~GeV/\textit{c} coupled to a low momentum track reconstruction cutoff (about 100 ${\rm MeV}/c$).
In addition, ALICE is equipped with detectors at forward and backward rapidities used for triggering and event characterisation purposes.

The main detectors used  for the collection of Run 2 data for this analysis are the ITS, the TPC, the Time Of Flight (TOF) detector~\cite{TOFtdr}, and the V0 detector~\cite{V0Performance}. 
A brief overview of the Run 2 configuration of each detector is provided below.

The ITS consists of six cylindrical layers of silicon detectors.
The two innermost layers are instrumented with Silicon Pixel Detectors (SPD), the two middle layers with Silicon Drift Detectors (SDD), and the outer layers with Silicon Strip Detectors (SSD).
The ITS provides a precise measurement of the location of the primary vertex (PV) of the collision and the properties of particles produced in the collisions, accurately pinpointing their path with a spatial resolution of around 5 micrometres.
The PV position can also be determined from track segments measured by the SPD layers (tracklets).

The TPC is the primary tracking detector of the ALICE experiment.
It consists of a large cylinder filled with a gas mixture, Ar/CO$_2$ (88/12) in 2016 and 2018, and Ne/CO$_2$/N$_2$ (90/10/5) in 2017, serving as a detection medium.
Multi-wire proportional chambers on the end plates collect the drift charge, providing spatial information of the track and pulse heights proportional to the ionisation strength of the charged particle.

The TOF detector covers the pseudorapidity range $|\eta|\lesssim0.9$ and the full azimuth with multi-gap resistive plate chambers (MRPCs).
Its primary purpose is the identification of particles with intermediate momentum via the measurement of their time of flight.
In this analysis, it is used to reduce the background from out-of-bunch pileup events, as discussed in Section~\ref{sec:analysis}.

The V0 detector consists of two arrays of scintillator counters, V0A and V0C, positioned on either sides of the interaction point.
V0C is placed upstream of the muon absorber at 0.9~m from the interaction point and covers the pseudorapidity range $\mathrm{-3.7 < \eta <-1.7}$.
V0A is located 3.3 meters away from the collision vertex on the opposite side and covers the pseudorapidity range $\mathrm{2.8 < \eta <5.1}$.
The V0 detector is used to estimate the collision multiplicity by summing up the charge deposited in the two disks, the total charge being correlated with the number of primary particles produced at midrapidity in the collision.
Additionally, the presence of a signal in both V0A and V0C provides the minimum-bias trigger.

The data used for this analysis consist of pp collisions at $\sqrt{s}$ = 13~TeV collected by ALICE from 2016 to 2018.
The events were selected using the high-multiplicity (HM) trigger, which is activated online when the signal amplitude in the V0 detectors exceeds a predefined threshold corresponding to approximately 30 charged particles produced at midrapidity per unit of rapidity. The HM events are selected in the multiplicity range 0--0.1\%, which includes the 0.1\% of the minimum-bias events characterised by the highest V0M amplitude, which is the sum of the amplitudes in the V0A and V0C detectors.

The criteria employed to select good-quality events are similar to those reported in Ref.~\cite{ALICE:2019avo}.
Only events with the primary vertex position along the beam axis within $\pm$10 cm around the nominal interaction point are considered.
Pileup events happening in the same bunch crossing are removed by excluding events with multiple vertices reconstructed with the SPD.
The separation along the z-coordinate between the PV reconstructed using tracklets, and the one built via tracks is required to be less than 5~mm when both PVs are available.
Based on the correlation between tracklets and clusters in the SPD, the background from beam-gas events is rejected offline.
After applying these criteria, the sample available for the analysis consists of around $4\times10^{8}$ events, corresponding to an integrated luminosity of approximately 
$7.7~\textrm{pb}^{-1}$.

\section{Analysis details}
\label{sec:analysis}
Due to their short lifetime, $\Xi^-$, $\overline{\Xi}^+$, $\Omega^-$, and $\overline{\Omega}^+$ baryons can only be detected indirectly and reconstructed from their decay products within the acceptance of the ALICE apparatus.
Their reconstruction exploits the following weak decay channels~\cite{ParticleDataGroup:2024cfk}:
\begin{gather*}
  \Xi^- (\, \overline{\Xi}^+ ) \rightarrow \Lambda (\overline{\Lambda}) + \pi^- (\pi^+) \hspace{1cm} \textrm{B.R.}\,\, 99.9 \% \hspace{1cm} c\tau = 4.91\,\textrm{cm}\,,
  \\
  \\
  \Omega^- (\,\overline{\Omega}^+) \rightarrow \Lambda (\overline{\Lambda}) + \textrm{K}^- (\textrm{K}^+) \hspace{1cm} \textrm{B.R.}\,\, 67.8 \% \hspace{1cm} c\tau = 2.461\,\textrm{cm}\,,
\end{gather*}
In the following, the sums of particles and antiparticles, $\Xi^- + \overline{\Xi}^+$ and $\Omega^- + \overline{\Omega}^+$, will be referred to as $\Xi$ and $\Omega$, respectively, and collectively referred to as multi-strange baryons.
The decay products $\pi$ and K are charged and can be tracked.
$\Lambda$ baryons are electrically neutral and need to be reconstructed from their decay products:
\begin{gather*}
  \Lambda (\overline{\Lambda}) \rightarrow \textrm{p} (\overline{\textrm{p}}) + \pi^- (\pi^+) \hspace{1cm} \textrm{B.R.}\,\, 64.1 \% \hspace{1cm} c\tau = 7.89\,\textrm{cm}\,.
\end{gather*}
The collected events are divided into three multiplicity classes: I, II, and III. These event classes are classified based on the V0M information computed from the sum of the amplitudes of the V0A and V0C detectors. The multiplicity classes correspond to three centrality intervals: 0.00--0.01\,\%, 0.01--0.05\,\%, and 0.05--0.10\,\%. Each multiplicity class is characterised by an average number of charged particles produced at midrapidity \avdndeta (hereafter referred to as \dndetashort), computed in Ref.~\cite{ALICE:2020swj}. The chosen multiplicity classes guarantee reasonable spacing in \dndetashort values and allow for the exploration of the highest multiplicities while ensuring sufficient statistics for reliable $\Omega$ measurements.

The identification of $\Xi$ and $\Omega$ relies heavily on the accuracy of the reconstructed daughter tracks.
For this reason, the track acceptance is limited to the pseudorapidity region $\lvert \eta \rvert < 0.8$ to guarantee uniform acceptance and performance.
Any track using shared ITS or TPC clusters (i.e., hits used by more than one track) is rejected, as it could be misassigned and worsen the tracking accuracy, as studied in Ref.~\cite{ALICE:2019avo}.
To ensure good momentum resolution and particle identification, tracks are required to cross at least 70 out of 159 readout pad rows of the TPC. Daughter particles are identified via a selection on the specific energy loss $\mathrm{d}E/\mathrm{d}x$ measured with the TPC detector, which is required to be compatible within 5$\sigma$ with the expected value, where $\sigma$ is the $\mathrm{d}E/\mathrm{d}x$ resolution.
The out-of-bunch pileup is discarded by requiring that at least one of the daughter tracks has a hit in the ITS or is matched to the TOF detector.
Both the ITS and the TOF detectors are significantly faster in readout than the TPC and, therefore, allow for the rejection of daughter particles that do not match a specific event. 

Geometric and kinematic selections are applied to the variables characterising decay topology.
In particular, selection criteria are defined based on those employed in previous studies~\cite{7TeV, ALICE:2019avo}.
An additional selection is applied to the pointing angle of the $\Lambda$ candidate built with the proton and the bachelor, i.e.,\ the $\pi$ (K) from the $\Xi$ ($\Omega$) decay. 
This angle, defined as the angle between the reconstructed momentum vector of the $\Lambda$ candidate and the line connecting the primary vertex to the $\Lambda$ decay vertex, is required to be larger than a predefined value, parametrised as a function of \pt, to reject bachelor-baryon pairs that are likely $\Lambda$ daughters.
In addition, a higher limit for the cascade pointing angle was used in order to improve the purity of the selected sample, from around 75\% (50\%) to over 90\% (80\%) for $\Xi$ ($\Omega$).

The multi-strange baryons are studied in the midrapidity window $\lvert y \rvert < 0.5$. In order to avoid contamination of $\Xi$ in the $\Omega$ sample, candidates are rejected if their invariant mass calculated under the $\Lambda \mathrm{\pi}$ assumption for the daughter particles lies within $\pm8$~MeV/$c^2$ from the nominal $\Xi$ mass.
The width of the rejection region is determined according to the invariant mass resolution of $\Xi$.
The candidates that have proper lifetimes larger than expected are also discarded.
All track, topological, and candidate selections applied to identify $\Xis$ and $\Omega$ are listed in Table~\ref{Tab_cuts}.

\begin{table}[htb!]
  \begin{center}
    \caption{Track, topological, and candidate selection criteria applied to identify $\Xi$ and $\Omega$ candidates.
      DCA stands for distance of closest approach.
      The symbol $\theta_\mathrm{P}$ stands for the pointing angle, i.e., the angle between the reconstructed momentum vector of the $\Lambda$ ($\Xi$) candidates and the line connecting the primary vertex to the $\Lambda$ ($\Xi$) decay vertex. The DCA between daughter tracks of the $\Lambda$ is measured in units of its resolution $\sigma$. More details about the selections are reported in the main text.
    }
    \renewcommand{\arraystretch}{1.2}
    \begin{tabular}{l|p{0.54\linewidth}}
      \hline \hline
      \multicolumn{2}{c}{\textbf{Track selections}}                                                                      \\
      \hline \hline
      Pseudorapidity                                           & $\lvert \eta \rvert  < 0.8$
      \\
      \hline
      {Number of TPC crossed pad rows $N_\mathrm{crossed}$}    & {$>$ 70}
      \\
      \hline
      TPC \dEdx                                                & $< 5\sigma$                                             \\
      \hline
      Out-of-bunch pileup rejection                            & At least one daughter track has a hit in the ITS or TOF \\
      \hline \hline
      \multicolumn{2}{c}{\textbf{Topological selections for $\Xi$ ($\Omega$)}}                                           \\
      \hline \hline
      Cascade transverse decay radius $R_\mathrm{min}$         & $> 0.6 (0.5)\, \textrm{cm}$                             \\
      \hline
      $\Lambda$ transverse decay radius $R_\mathrm{max}$       & $> 1.2 (1.1)\, \textrm{cm}$                             \\
      \hline
      DCA bachelor to PV                                       & $> 0.04\, \textrm{cm}$                                  \\
      \hline
      DCA $\Lambda$ to PV                                      & $> 0.06\, \textrm{cm}$                                  \\
      \hline
      DCA meson daughter to PV                                 & $> 0.04\, \textrm{cm}$
      \\
      \hline
      DCA baryon daughter to PV                                & $> 0.03\, \textrm{cm}$                                  \\
      \hline
      DCA between daughter tracks of the $\Lambda$             & $< 1.5\sigma$                                           \\
      \hline
      DCA between the bachelor and the $\Lambda$               & $< 1.3\, \textrm{cm}$                                   \\
      \hline
      {$\mathrm{cos(\theta_P)}$ (of cascade to PV)}            & {$>$ 0.998}                                             \\
      \hline
      {$\mathrm{cos(\theta_P)}$ (of $\Lambda$ to PV)}          & {$>$ 0.97}                                              \\
      \hline
      $\Lambda$ mass window                                    & $\pm 8$ MeV/$c^2$
      \\
      \hline
      $\mathrm{cos(\theta_P)}$ (of bachelor-baryon pair to PV) & \pt-dependent selection                                 \\
      \hline \hline
      \multicolumn{2}{c}{\textbf{Candidate selections}}                                                                  \\
      \hline \hline
      Rapidity interval                                        & $\lvert y \rvert  < 0.5$                                \\
      \hline
      Proper lifetime $\tau$                                   & $< 3 \times \langle\tau\rangle$                         \\
      \hline
      Competing cascade rejection (only for $\Omega$)          & $\lvert M(\Xi) - 1.321 \rvert > 8$ MeV/$c^2$            \\
      \hline  \hline
    \end{tabular}
    \label{Tab_cuts}
  \end{center}
\end{table}

The invariant mass distributions of the candidates passing the aforementioned selections are fitted with the sum of a Gaussian describing the signal and a linear function describing the residual background~\cite{ALICE:2019avo}.
The signal is extracted by subtracting the integral of the background region from the integral of the invariant mass distributions in the interval ($\mu - 4\sigma_G$, $\mu + 4\sigma_G$), where $\mu$ and $\sigma_G$ are the mean and width of the Gaussian fit function, respectively. This procedure was used in order not to introduce any systematic uncertainty associated to the fit description of the invariant mass peak.

The acceptance and efficiency of reconstructed multi-strange baryons are computed using a Monte Carlo (MC) simulation with event generation based on \textsc{Pythia}8.2 with the Monash 2013 tune~\cite{Sjostrand:2014zea}, and particle propagation through the detector described with {\textsc{Geant}~3}~\cite{Brun:1073159}.
The simulation reproduces the data-taking conditions.
The efficiency shows a negligible dependence on multiplicity. Therefore, the efficiency computed in the 0-100\% multiplicity class is used in all multiplicity intervals.

A L\'{e}vy-Tsallis function~\cite{Tsallis:1987eu} is used to fit the $p_{\rm T}$ spectra in order to extrapolate the yield in the unmeasured \pt interval.
The extrapolated fraction of the $\Xi$ yield is approximately 15\%, whereas for $\Omega$ yield it is approximately 25\%.

\subsection{Systematic uncertainties}
Candidate selection criteria, material budget, signal extraction, and in-bunch pileup are the four major contributors to the total systematic uncertainty. Uncertainties are estimated in the 0-0.1\% V0M class and are assigned to all three multiplicity classes considered in this analysis.

The systematic uncertainty associated with the topological and track selections is computed by varying the selection criteria around the default values and then comparing the results with the default ones, as done in previous analyses and reported in detail in Ref.~\cite{ALICE:2019avo}.
The variations of the selection criteria are the same as those used in Ref.~\cite{ALICE:2019avo}, and each of them determines a change in the raw signal smaller than 10\%.
The criterion proposed by R. Barlow~\cite{Barlow:2002yb} is applied to assess whether a systematic variation is statistically significant: variations not compatible with statistical fluctuations are added to the systematic uncertainty.
This uncertainty decreases with \pt, varying for \Xis (\Oms) from 5\% (12\%) at $\pt\approx1$~GeV/\textit{c} to 2\% (3\%) for \pt$>$ 2 GeV/\textit{c}.

The systematic uncertainty associated with the TPC d$E$/d$x$ selection was assessed by varying the selection criteria between 4$\sigma$ and 7$\sigma$, resulting in a maximum deviation of the yields of 0.3\% for $\Xi$ and 1\% for $\Omega$. The uncertainty related to competing decay rejection was evaluated by entirely removing this selection for $\Omega$, leading to a maximum deviation of 2\% in the $p_{\text{T}}$ spectra. The uncertainty due to the proper lifetime ($\tau$) selection, varied between 2.5 and 5~$\tau$, was found to be 0.5\% for $\Xi$ and 1\% for $\Omega$.

The impact of out-of-bunch pileup rejection on the systematic uncertainty was assessed by modifying the matching scheme with the relevant detectors, considering both no matching and ITS matching of at least one decay track over the full $p_{\text{T}}$ range. This uncertainty was found to increase with transverse momentum and saturate at high $p_{\text{T}}$, reaching a maximum of 2\% for both $\Xi$ and $\Omega$.

The uncertainty given by the signal extraction procedure is tested by employing two different variations.
First, different ranges for signal extraction were considered.
The invariant mass interval around the mean value $\mu$ is loosened to $\pm 5 \sigma$ ($\pm 4.5 \sigma$) and tightened to $\pm 3 \sigma$ ($\pm 3.5 \sigma$) for $\Xi$ ($\Omega$).
The uncertainty associated with this source is approximately 0.5\% (1\%) for $\Xi$ ($\Omega$).
Second, two different background functions are used: a second- and third-degree polynomial.
The corresponding systematic uncertainty is found to be smaller than 0.3\%.

To account for any dependence of the efficiency on the multiplicity, a relative systematic uncertainty of 2\%, correlated across \pt intervals, is assigned to the spectra, as done in Ref.~\cite{ALICE:2019avo}.
The inaccuracies in the description of the material of the ALICE apparatus are accounted for in the material budget uncertainty.
This is estimated using a MC simulation in which the amount of material in the ALICE detector is varied within its uncertainty range ($\pm 4.5\%$).
The systematic uncertainty from this source is below 5\% for $\Xi$ and below 3\% for $\Omega$ in the lowest $p_{\textrm T}$ interval, and less than 2\% for higher $p_{\textrm T}$ values.
Finally, the uncertainty related to the in-bunch pileup contribution is estimated to be smaller than 2\% in all \pt intervals~\cite{ALICE:2019avo}.
A summary of the relative systematic uncertainties of the $\Xi$ and $\Oms$ spectra is listed in Table~\ref{tab:sysPtSpectra}.

To extrapolate the systematic uncertainty of the \pt spectra to the unmeasured \pt interval, the $p_{\textrm T}$ spectra are increased or decreased by $1\sigma$ of systematic uncertainty and fitted with the L\'{e}vy-Tsallis function.
The half-difference between the obtained maximum and the minimum \pt-integrated yields provides the systematic uncertainty of the yield.
A similar procedure is applied to compute the systematic uncertainty of the \meanpt. In this case, assuming the systematic uncertainties are uncorrelated with \pt, each point in the $p_{\textrm T}$ spectra is increased or decreased within $1\sigma$ of systematic uncertainty in order to get the hardest and softest spectra, i.e.,\ the spectra that maximise and minimise the \meanpt, respectively.
The half-difference between the maximum and the minimum \meanpt provides the systematic uncertainty of the \meanpt.
The other contribution is related to the extrapolated fraction of the fitted spectra.
The spectra are fitted with four alternative functions: Boltzmann, Bose-Einstein, m$_\textrm{T}$-exponential, and Fermi-Dirac~\cite{ALICE:2019avo}.
The \pt-integrated yields and \meanpt are compared to the default ones obtained with the L\'{e}vy-Tsallis function, and the maximum half-difference is assigned as a systematic uncertainty.

\begin{table}[!htbp]
  \caption{Summary of the relative systematic uncertainties of the \Xis (top) and the $\Omega$ (bottom) \pt spectra measured in the V0M multiplicity class 0-0.1\%. The contribution labelled ``Candidate selection criteria’’ refers to systematic
    uncertainties related to selections listed in Table~\ref{Tab_cuts}. }
  \begin{center}
    \renewcommand{\arraystretch}{1.35}
    \resizebox{\textwidth}{!}{
      \begin{tabular}{|m{0.4\textwidth}|>{\centering}m{0.2\textwidth}>{\centering}m{0.2\textwidth}>{\centering\arraybackslash}m{0.2\textwidth}|}
        \hline
        \multicolumn{1}{|c|}{{\textbf{Hadron}}}               &
        \multicolumn{3}{c|}{{$\Xi$}}                                                                                   \\
        \multicolumn{1}{|c|}{$p_\mathrm{T}$ (GeV/\textit{c})} &
        $\approx$1.0                                          & $\approx$2.0     & $\approx$3.5                        \\
        \hline
        \hline
        {Candidate selection criteria}                        & {5\%}            & {2\%}            & {2\%}            \\
        \hline
        {Signal extraction window}                            & {0.5\%}          & {0.5\%}          & {0.5\%}          \\
        \hline
        {Background fit function}                             & {$<$0.3\%}       & {$<$0.3\%}       & {$<$0.3\%}       \\
        \hline
        {Multiplicity dependence of efficiency}               & {2\%}            & {2\%}            & {2\%}            \\
        \hline
        {Material budget}                                     & {5\%}            & {2\%}            & {1\%}            \\
        \hline
        {Residual in-bunch pileup}                            & {2\%}            & {2\%}            & {2\%}            \\
        \hline
        {\textbf{Total}}                                      & \textbf{{7.5\%}} & \textbf{{4\%}}   & \textbf{{3.5\%}} \\
        \hline
        \hline
        \multicolumn{1}{|c|}{{\textbf{Hadron}}}               &
        \multicolumn{3}{c|}{{$\Omega$}}                                                                                \\
        \multicolumn{1}{|c|}{$p_\mathrm{T}$ (GeV/\textit{c})} &
        $\approx$1.0                                          & $\approx$2.0     & $\approx$3.5                        \\
        \hline
        {Candidate selection criteria}                        & {12\%}           & {3\%}            & {3\%}            \\
        \hline
        {Signal extraction window}                            & {1\%}            & {1\%}            & {1\%}            \\
        \hline
        {Background fit function}                             & {$<$0.3\%}       & {$<$0.3\%}       & {$<$0.3\%}       \\
        \hline
        {Multiplicity dependence of efficiency}               & {2\%}            & {2\%}            & {2\%}            \\
        \hline
        {Material budget}                                     & {3\%}            & {1.5\%}          & {1.5\%}          \\
        \hline
        {Residual in-bunch pileup}                            & {2\%}            & {2\%}            & {2\%}            \\
        \hline
        {\textbf{Total}}                                      & \textbf{{13\%}}  & \textbf{{4.5\%}} & \textbf{{4.5\%}} \\
        \hline
      \end{tabular}
    }
    \label{tab:sysPtSpectra}
  \end{center}
\end{table}

\section{Results and discussion}
\label{sec:results}
\begin{figure}[htb!]
  \centering
  \begin{minipage}{0.45\textwidth}
    \hskip -20mm
    \centering
    \includegraphics[width=1.1\textwidth]{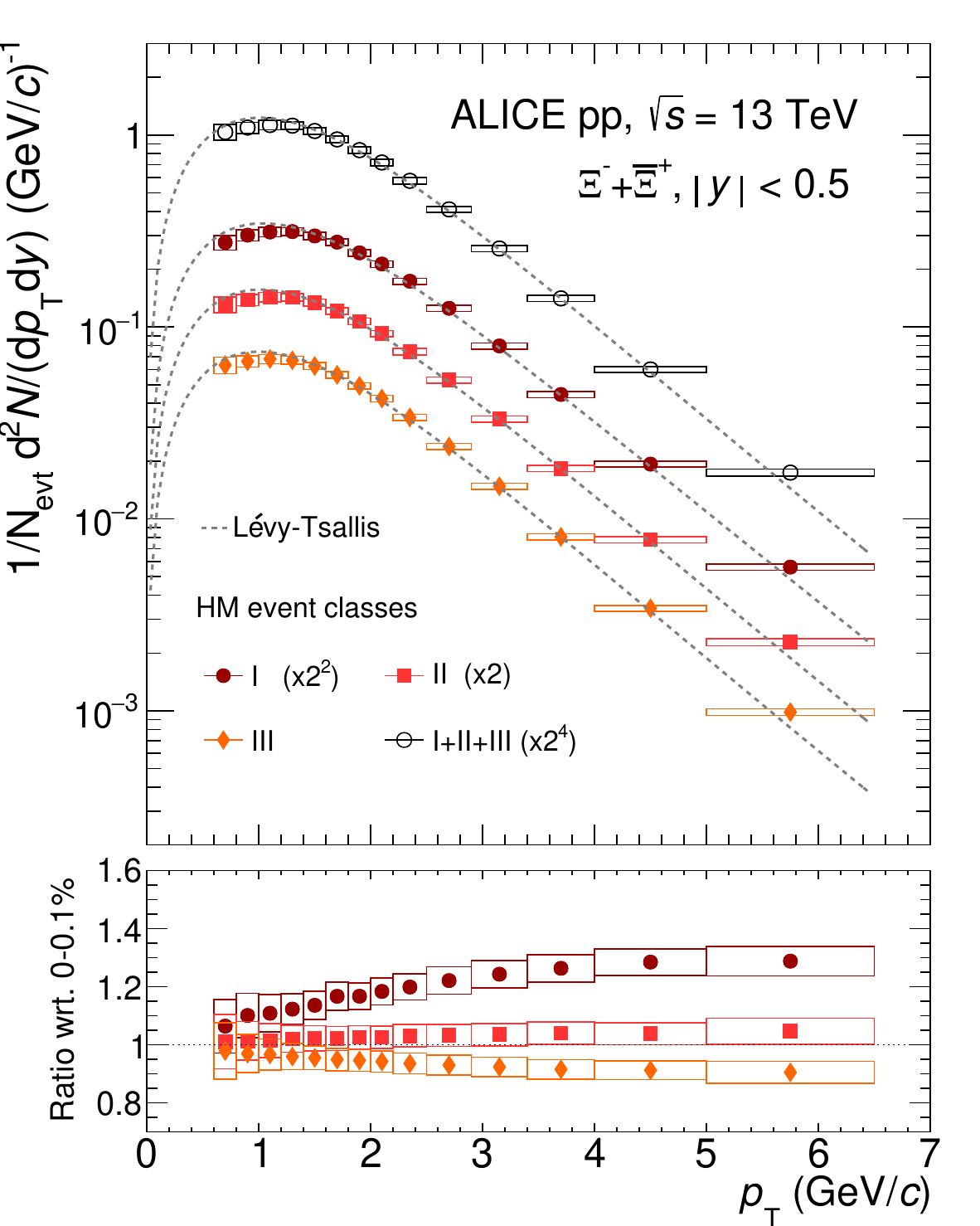}
  \end{minipage}
  \begin{minipage}{0.45\textwidth}
    \centering    \includegraphics[width=1.1\textwidth]{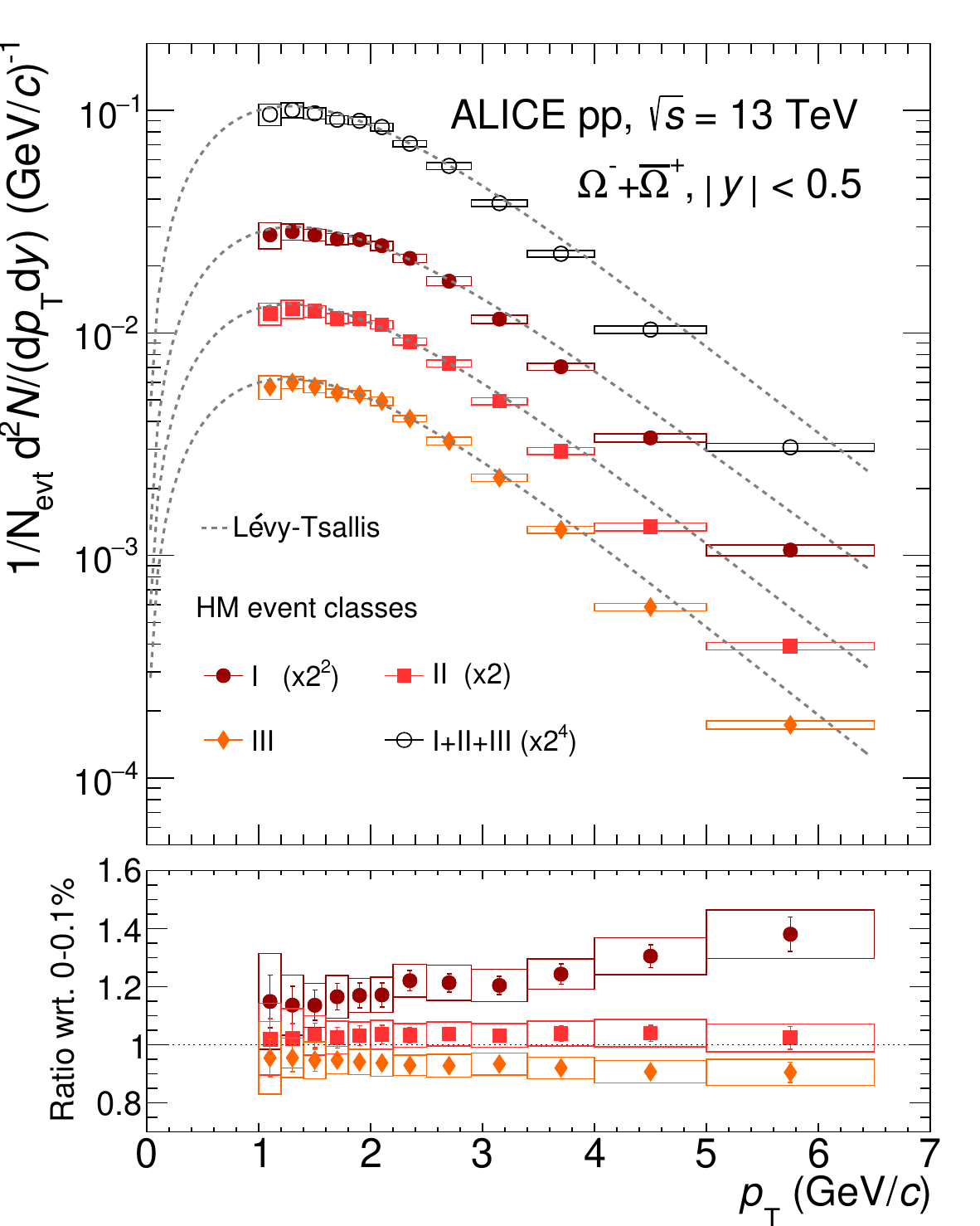}
  \end{minipage}
  \caption{$\Xi$ (left) and $\Omega$ (right) $\pt$ spectra fitted with the L\'{e}vy-Tsallis function. Different colours refer to different multiplicity classes. The spectra are scaled by different factors to improve the visibility. The bottom panels display the ratios to the spectrum measured in the 0--0.1\% multiplicity class. Error bars and boxes represent statistical and systematic uncertainties, respectively.}
  \label{LevyFit}
\end{figure}

The corrected $\Xi$ and $\Omega$ $p_{\textrm T}$ spectra measured in the three different multiplicity classes considered in the analysis and in the 0-0.1\% V0M class are shown in the left and right top panels of Fig.~\ref{LevyFit}, respectively. In the plots, the L\'{e}vy-Tsallis fit functions are displayed with dashed lines.
In the bottom panels, the ratios
to the spectra measured in the 0--0.1\% multiplicity class are shown.
The yields increase with multiplicity in all \pt intervals, and the spectra become harder as the multiplicity increases.
This behaviour was already observed in previous measurements performed with a minimum-bias sample of pp collisions~\cite{ALICE:2019avo}.

\begin{figure}[htb!]
  \centering
  \begin{minipage}{0.45\textwidth}
    \hskip -10mm
     \includegraphics[width=1.1\textwidth]{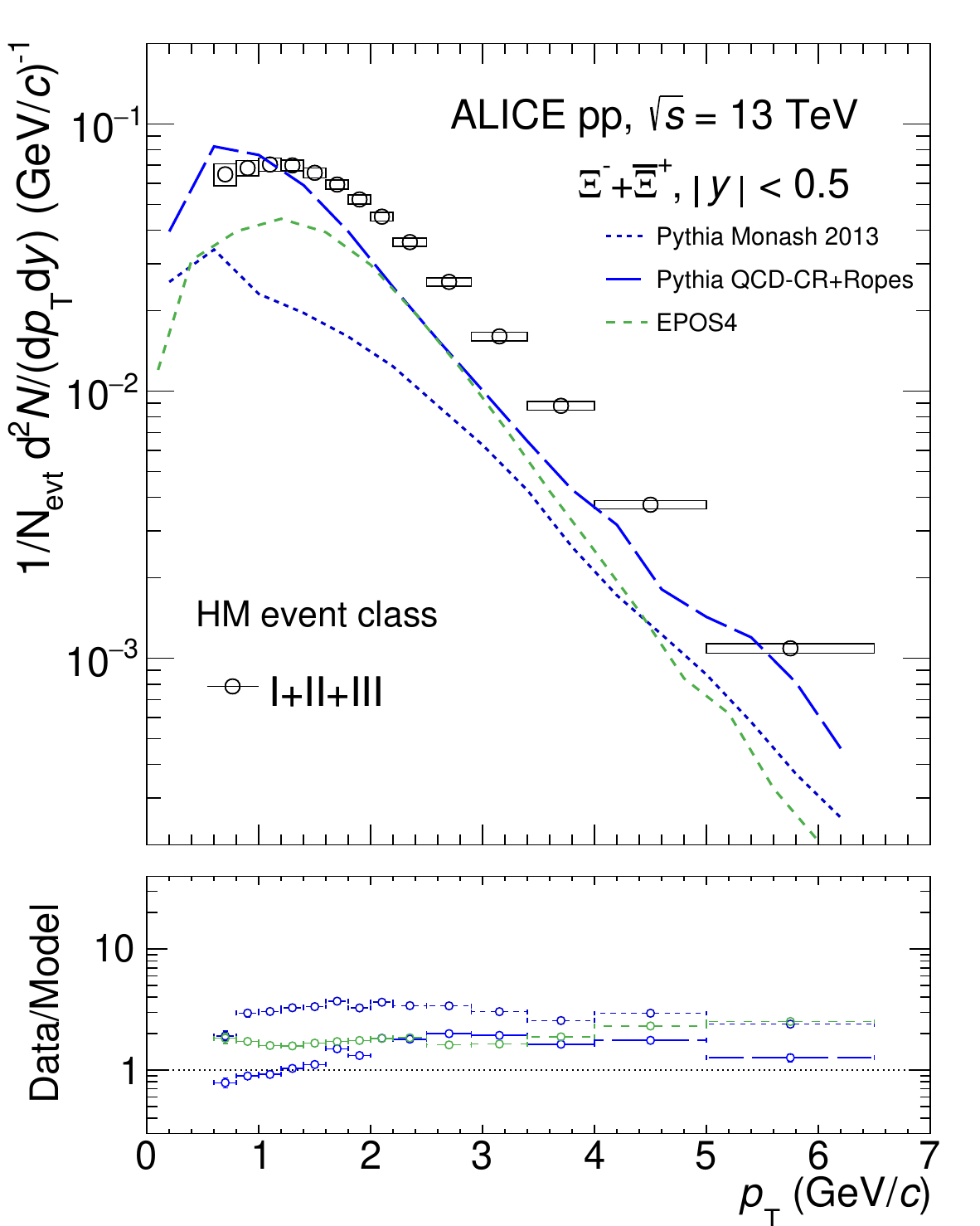}
  \end{minipage}
  \begin{minipage}{0.45\textwidth}
    \centering
     \includegraphics[width=1.1\textwidth]{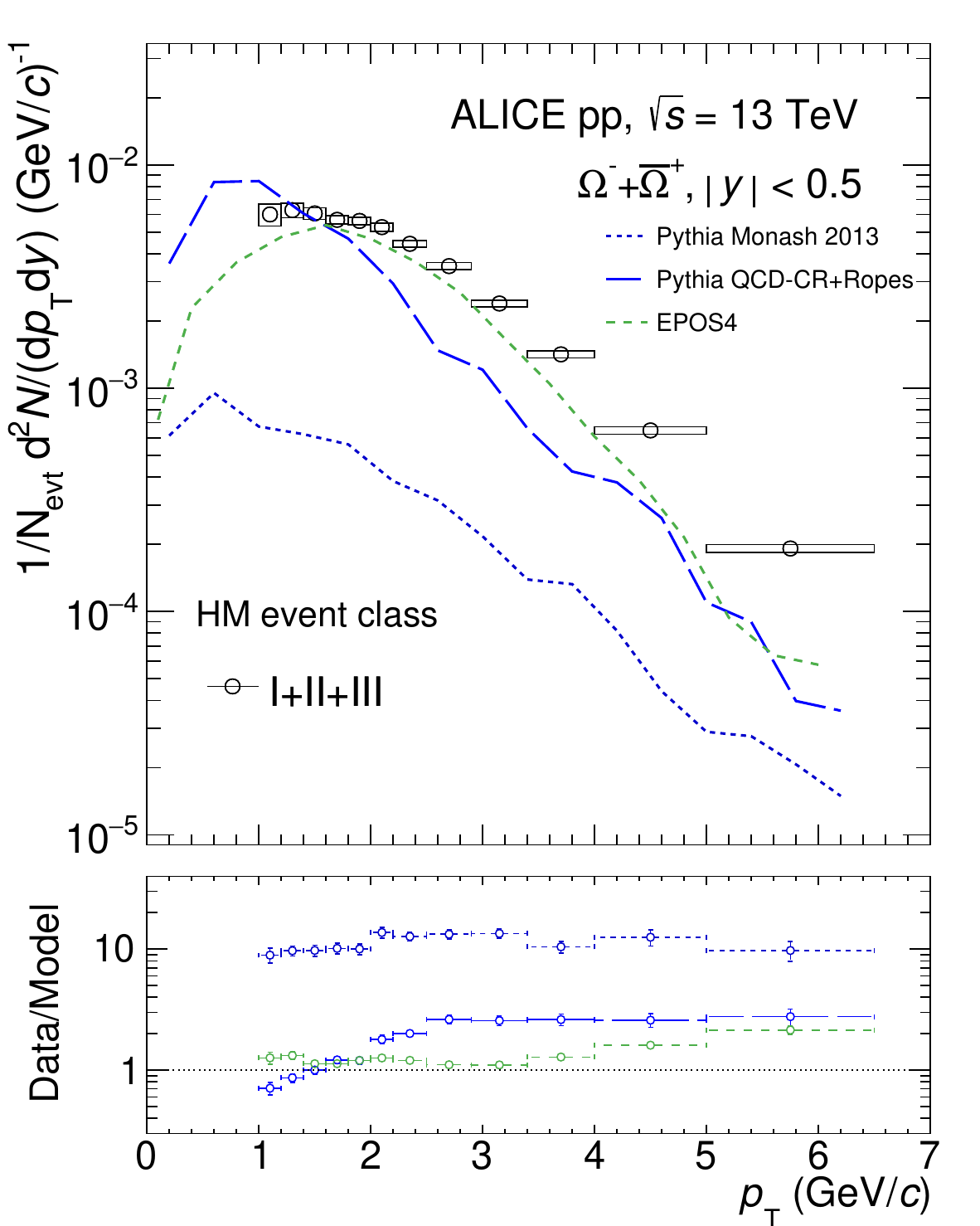}
  \end{minipage}
  \caption{
    $\Xi$ (left) and $\Omega$ (right) $\pt$ spectra in the multiplicity class 0-0.1\% compared to predictions from different models. The data points are shown with markers, and the model predictions~\cite{Sjostrand:2014zea, Bierlich:2014xba, EPOS4_PhysRevC.108.034904, EPOS4_PhysRevC.108.064903, EPOS4_PhysRevC.109.014910} with lines of different styles. Error bars and boxes represent statistical and systematic uncertainties, respectively.
    The bottom panels show the ratios between the spectra and the model, with error bars representing total uncertainties obtained by propagating the uncertainties of both the data and the model.
  }
  \label{fig:models}
\end{figure}

In Fig.~\ref{fig:models}, the \pt spectra of $\Xi$ (left panel) and $\Omega$ (right panel) for the multiplicity class 0-0.1\% are compared with the predictions of three different phenomenological models, namely \textsc{Pythia}8.2 with the Monash 2013 tune~\cite{Sjostrand:2014zea}, \textsc{Pythia}8.2 with Ropes~\cite{Bierlich:2014xba} and EPOS4~\cite{EPOS4_PhysRevC.108.034904, EPOS4_PhysRevC.108.064903, EPOS4_PhysRevC.109.014910}. The bottom panels show the ratios of the spectra to the integral of the model predictions in each \pt interval. The error bars represent the total uncertainties obtained by propagating the uncertainties of both the data and the model.
\textsc{Pythia} is based on the Lund string hadronisation model~\cite{Andersson:1997xwk}.
\textsc{Pythia}8.2 with the Monash 2013 tune implements Multi Parton Interactions (MPI) and a basic version of the Colour Reconnection (CR) mechanism.
\textsc{Pythia}8.2 with Ropes allows overlapping strings to interact with each other, forming the so-called ``colour ropes"~\cite{Bierlich:2014xba}.
As shown in Ref.~\cite{Pion}, \textsc{Pythia}8.2 with Ropes improves the description of strange hadron production in pp collisions with respect to \textsc{Pythia}8.2 with the Monash 2013 tune.
Finally, the EPOS4 event generator is the newest version of the EPOS model. This model implements the core-corona approach~\cite{PhysRevLett.98.152301}, according to which strings in a low-density area (``corona'') hadronise via string fragmentation, while in a high-density area they fuse into a ``core'' which expands hydrodynamically and undergoes statistical hadronisation.
The multiplicity classes in the MC are defined starting from the multiplicity of generated particles in the V0 acceptance.
The comparison with the data is carried out in the corresponding percentile classes.
As such, the predictions for the $p_{\rm T}$ distribution depend also on the model's capability to reproduce the forward/midrapidity correlations. While the models capture the overall trends of the data, some quantitative differences remain, as the predicted multiplicities are underestimated by about 10\%.
The comparison between the data and the model predictions shows that
\textsc{Pythia}8.2 with the Monash 2013 tune underestimates the $\Xi$ yields by about 50 to 70\%, depending on the \pt interval, and the \Oms yield by about 90\%.
\textsc{Pythia}8.2 with Ropes  improves the spectra description, especially at low \pt, where it overestimates the yield by about 20\%. On the contrary, for $3 < p_\mathrm{T} < 5$~GeV/\textit{c}, it underestimates the $\Xi$ ($\Omega$) spectra by approximately 40\% (70\%), thus predicting softer spectra than the measured ones.
A better description of the $\Omega$ spectral shape is provided by EPOS4, which underestimates the yield by less than 20\% in the $1 < p_\mathrm{T} < 4$~GeV/\textit{c} interval. Interestingly, the EPOS4 description of $\Xi$ at intermediate \pt is worse than for $\Omega$, as it underestimates the yield by approximately 40\% for $p_\mathrm{T} < 3$~GeV/\textit{c}. Both particle spectra are underestimated by about 70\% for $p_\mathrm{T} > 5$~GeV/\textit{c}.

\begin{figure}[htb!]
  \centering
  \begin{minipage}{0.45\textwidth}
    \centering
    \hskip -10mm
\includegraphics[width=1.1\textwidth]{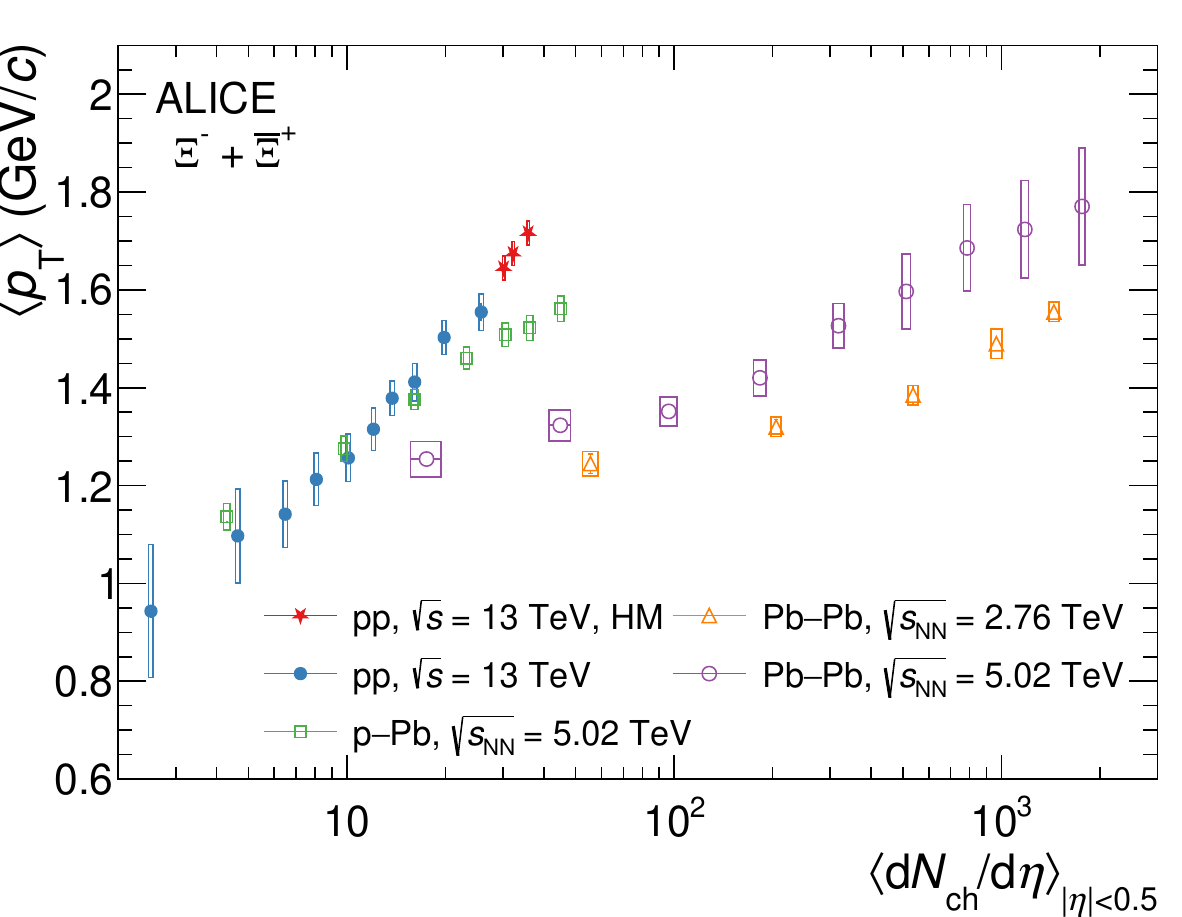}
  \end{minipage}
  \begin{minipage}{0.45\textwidth}
    \centering
\includegraphics[width=1.1\textwidth]{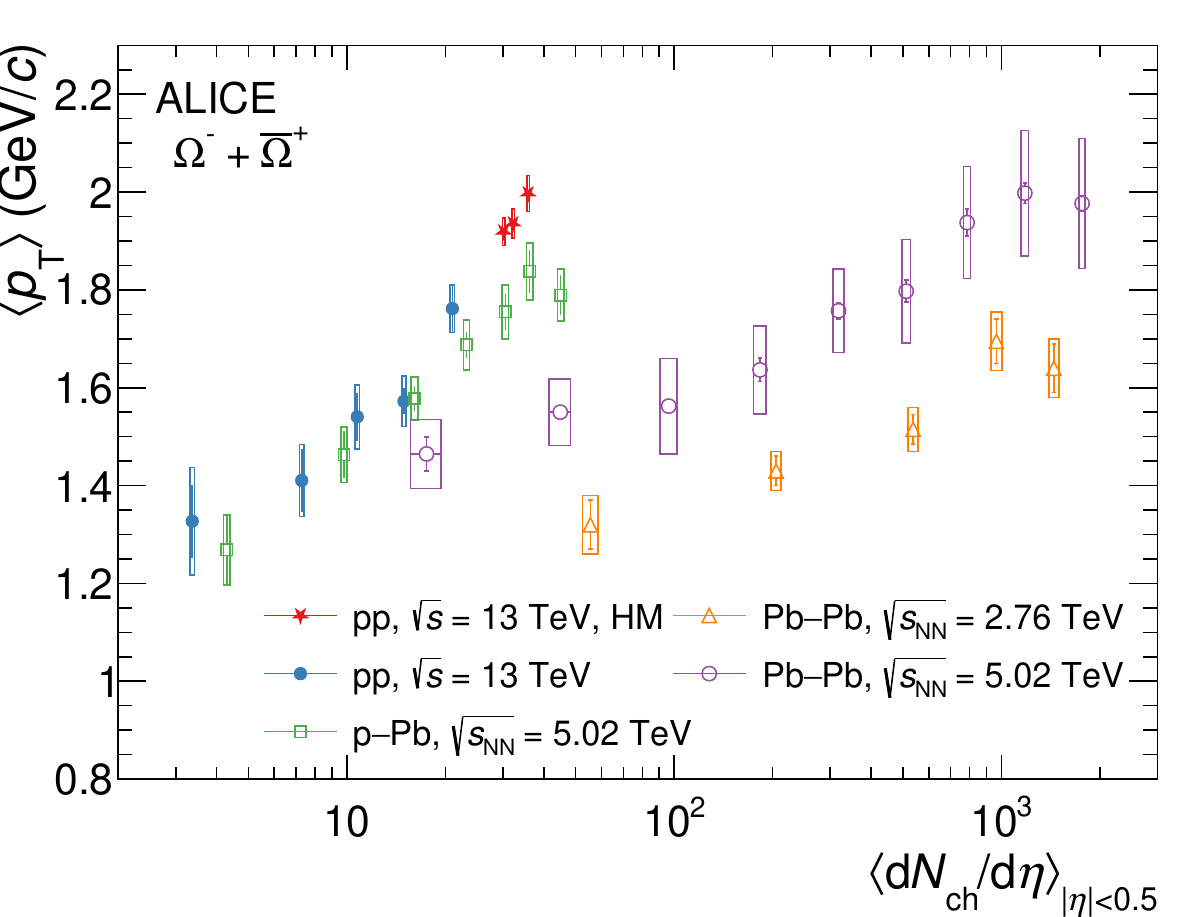}
  \end{minipage}
  \caption{
    $\Xi$ (left) and $\Omega$ (right) mean $\pt$ (\meanpt) as a function of charged-particle multiplicity measured at midrapidity in pp, p--Pb and Pb--Pb collisions, as indicated in the legend~\cite{ALICE:2019avo, pPb5TeV, PbPb276TeV, PbPb5TeV}. The red markers show the results in the three HM classes presented in this article. Error bars and boxes represent statistical and systematic uncertainties, respectively.
  }
  \label{meanpt}
\end{figure}

The left and right panels of Fig.~\ref{meanpt} display the mean \pt of $\Xi$ and $\Omega$, respectively, as a function of the charged-particle multiplicity measured at midrapidity. The results obtained in HM pp collisions (red markers) are compared to previous results in pp, p--Pb and Pb--Pb collisions~\cite{ALICE:2019avo, pPb5TeV, PbPb276TeV, PbPb5TeV}. The \meanpt measured in HM pp collisions is larger than the one measured in p--Pb and Pb--Pb collisions at similar multiplicities and follows the same trend shown by previous results in pp collisions at $\sqrt{s}=$13~TeV~\cite{ALICE:2019avo}. While previous measurements showed only a $2\sigma$ difference between pp and p--Pb results at $\dndetashort\sim20$, the difference between the HM pp points and the p--Pb results at similar multiplicities amounts to at least $6\sigma$ for $\Xi$ and $4\sigma$ for $\Omega$, thus highlighting a clear divergence in trends between the two systems. The observed difference between the trends of the pp, p--Pb and Pb--Pb data in the overlapping multiplicity region (corresponding to high-multiplicity pp and p--Pb collisions and low-multiplicity Pb--Pb collisions) was already reported for charged particles~\cite{ALICE:2013rdo} and strange hadrons~\cite{PbPb5TeV}, and suggests that similar charged-particle multiplicities are obtained via harder interactions as the collision system becomes smaller.

\begin{figure}[htb!]
  \centering
  \begin{minipage}{0.45\textwidth}
    \centering
    \hskip -10mm
    \includegraphics[width=1.1\textwidth]{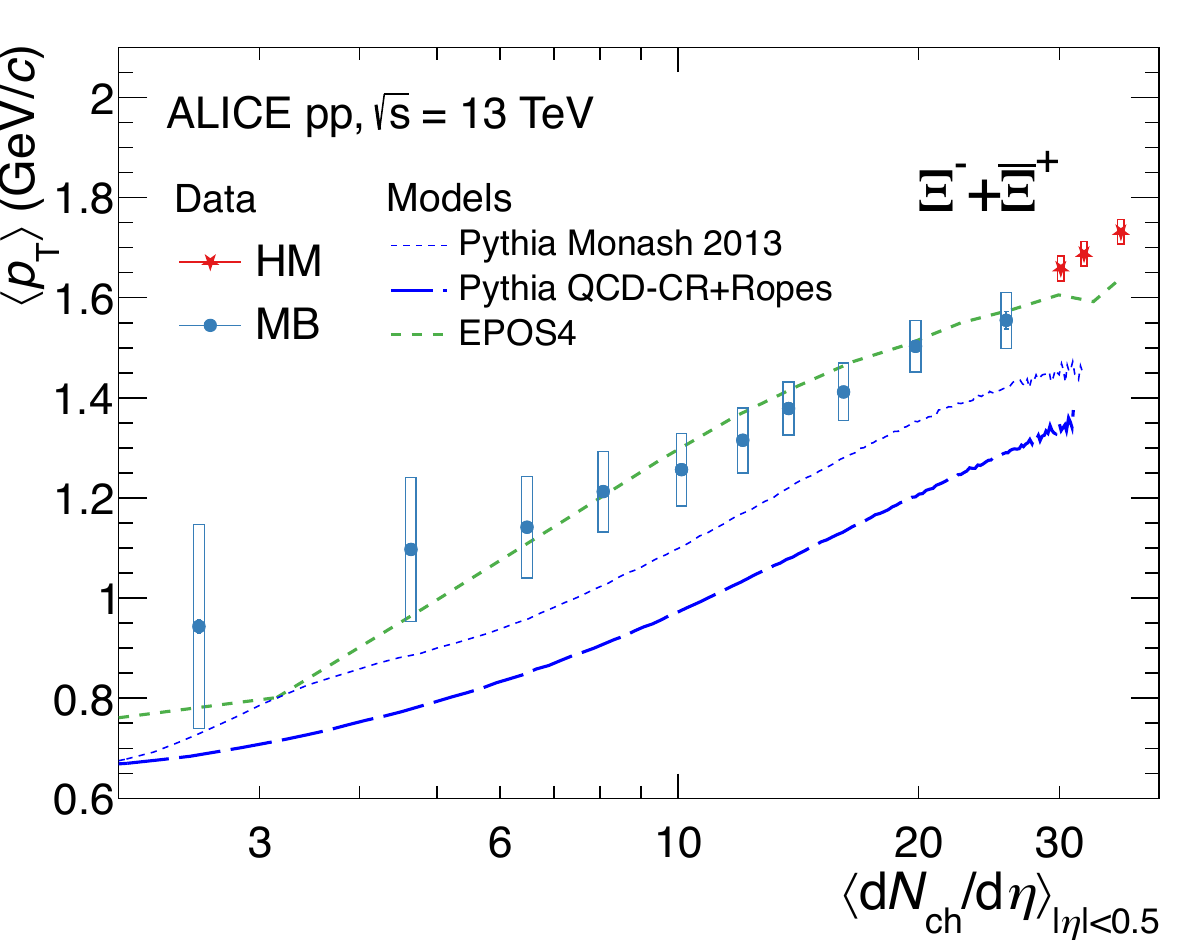}
  \end{minipage}
  \begin{minipage}{0.45\textwidth}
    \centering
    \includegraphics[width=1.1\textwidth]{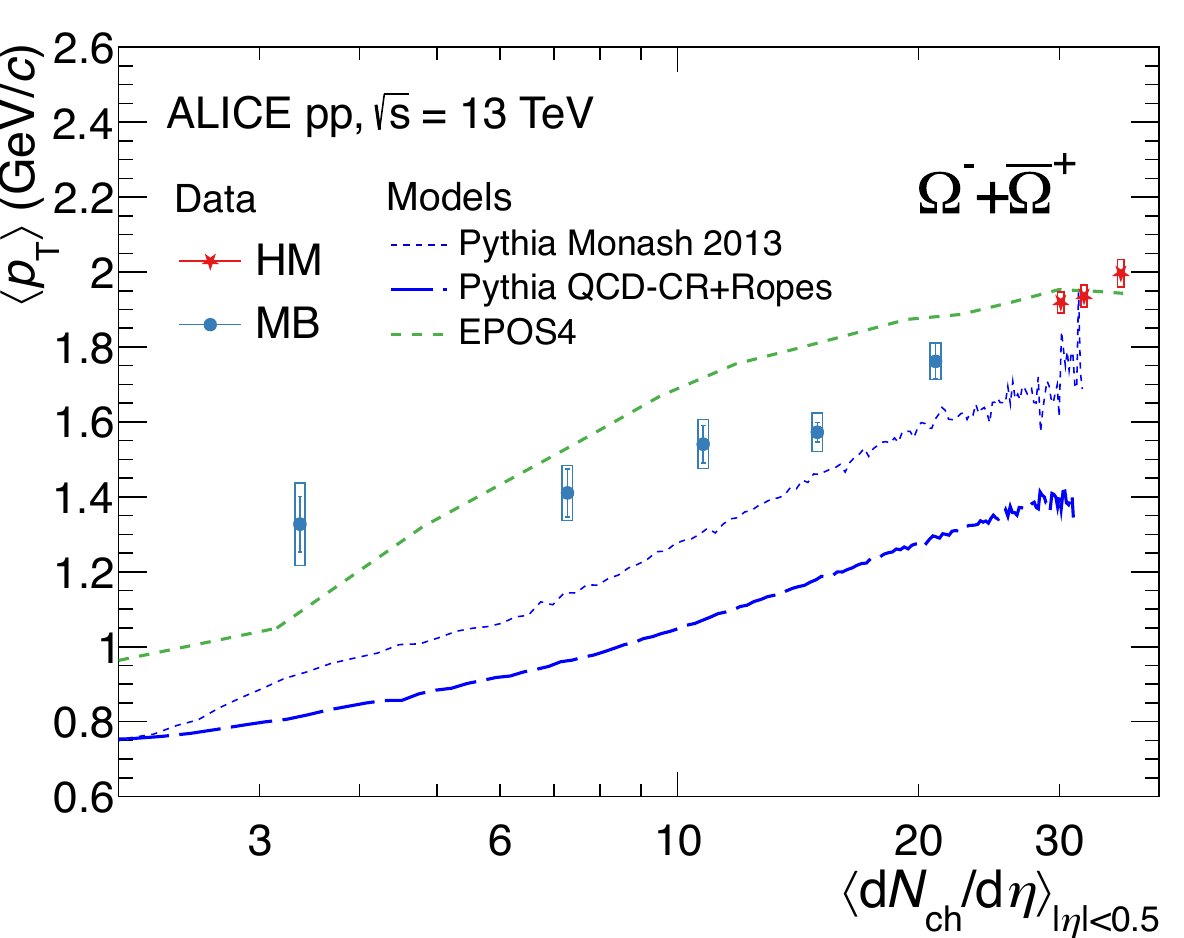}
  \end{minipage}
  \caption{$\Xi$ (left) and $\Omega$ (right) mean $\pt$ (\meanpt) as a function of charged-particle multiplicity measured at midrapidity in pp collisions~\cite{ALICE:2019avo}, compared to models~\cite{Sjostrand:2014zea, Bierlich:2014xba, EPOS4_PhysRevC.108.034904, EPOS4_PhysRevC.108.064903, EPOS4_PhysRevC.109.014910}. The red markers show the results in the three HM classes presented in this article. Error bars and boxes represent statistical and systematic uncertainties, respectively.}
  \label{meanpt_model}
\end{figure}

The \meanpt values measured in pp collisions at $\sqrt{s}=13$~TeV are compared to the predictions of \textsc{Pythia}8.2 with the Monash 2013 tune, \textsc{Pythia}8.2 with Ropes and EPOS4 in the left and right panels of Fig.~\ref{meanpt_model} for \Xis and \Oms, respectively. In the models, \meanpt is computed as the average of the simulated \pt spectrum in the full \pt range.
All models reproduce the increasing trend of \meanpt with charged-particle multiplicity. Although ropes in PYTHIA are needed to describe the enhanced production of multi-strange baryons at low \pt (see Fig.~\ref{fig:models}), the effect does not improve the \meanpt dependence as a function of multiplicity. On the contrary, the addition of ropes in \textsc{Pythia}8.2  increases the underestimation of \meanpt. 
The \meanpt of \Xis predicted by EPOS4 is consistent within less than $1.5\sigma$ with the data for \dndetashort $\lesssim30$, whereas the \meanpt values measured in the two highest multiplicity classes are slightly larger than the model prediction. The \meanpt of \Oms is well reproduced by EPOS4 in the three highest multiplicity classes, but is slightly overestimated at intermediate multiplicity ($10 \lesssim \dndetashort \lesssim 20$).
\vskip -2mm

\begin{figure}[!htbp]
  \centering
  \includegraphics[width=0.9\textwidth]{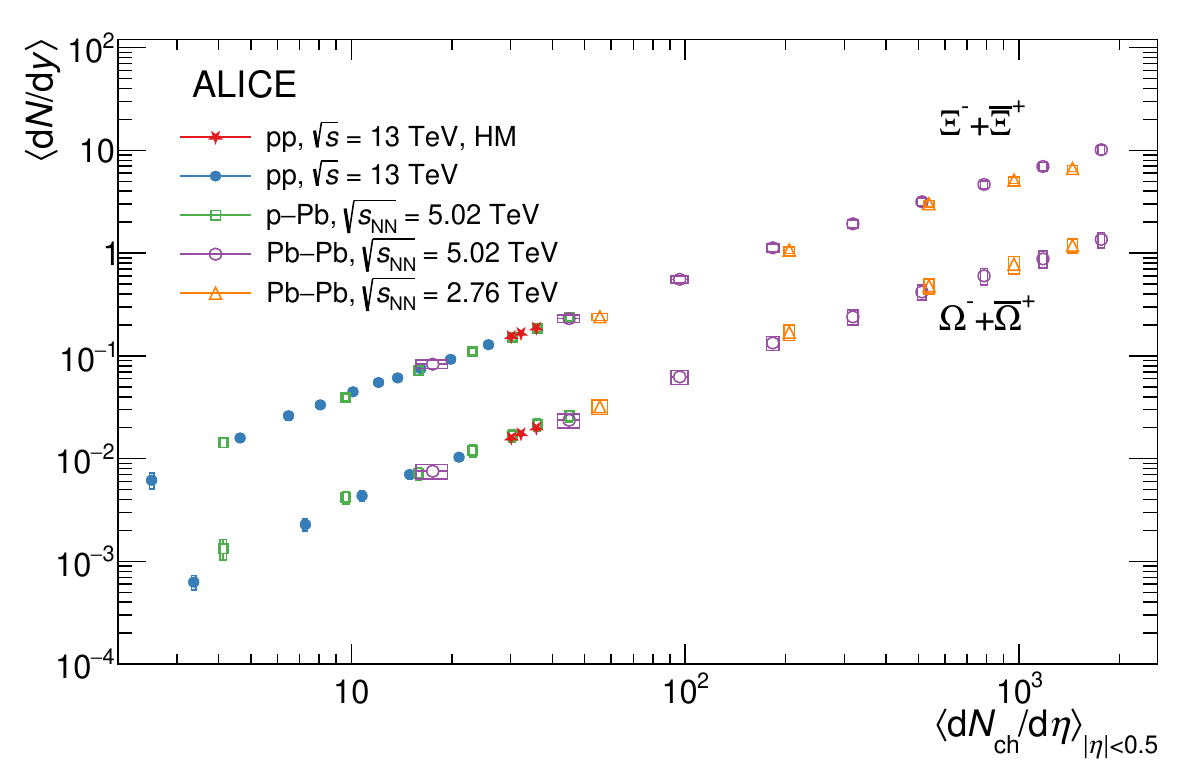}
  \vskip -2mm
  \caption{$\Xi$ and $\Omega$ $\pt$-integrated yields per event as a function of charged-particle multiplicity measured at midrapidity in pp, p--Pb and Pb--Pb collisions, as indicated in the legend~\cite{ALICE:2019avo, pPb5TeV, PbPb276TeV, PbPb5TeV}. The red markers show the results in the three HM classes presented in this article. Error bars and boxes represent statistical and systematic uncertainties, respectively.
  }
  \label{int_yield_pbpb}
\end{figure}

The dependence of the $\pt$-integrated yields on the charged-particle multiplicity is shown in Fig.~\ref{int_yield_pbpb}. The yields are compared to those measured in other systems, namely p--Pb at 5.02~TeV~\cite{pPb5TeV}, and Pb--Pb at 2.76~TeV~\cite{PbPb276TeV} and 5.02~TeV~\cite{PbPb5TeV}.
The yields increase smoothly with multiplicity across collision systems; the yields in pp, p--Pb, and Pb--Pb collisions are similar at a given multiplicity. This observation indicates that the number of produced $\Xi$ and $\Omega$ is strongly correlated with multiplicity, regardless of the system size and energy, at the LHC energies.

\begin{figure}[!ht]
  \centering
  \includegraphics[width=0.9\textwidth]{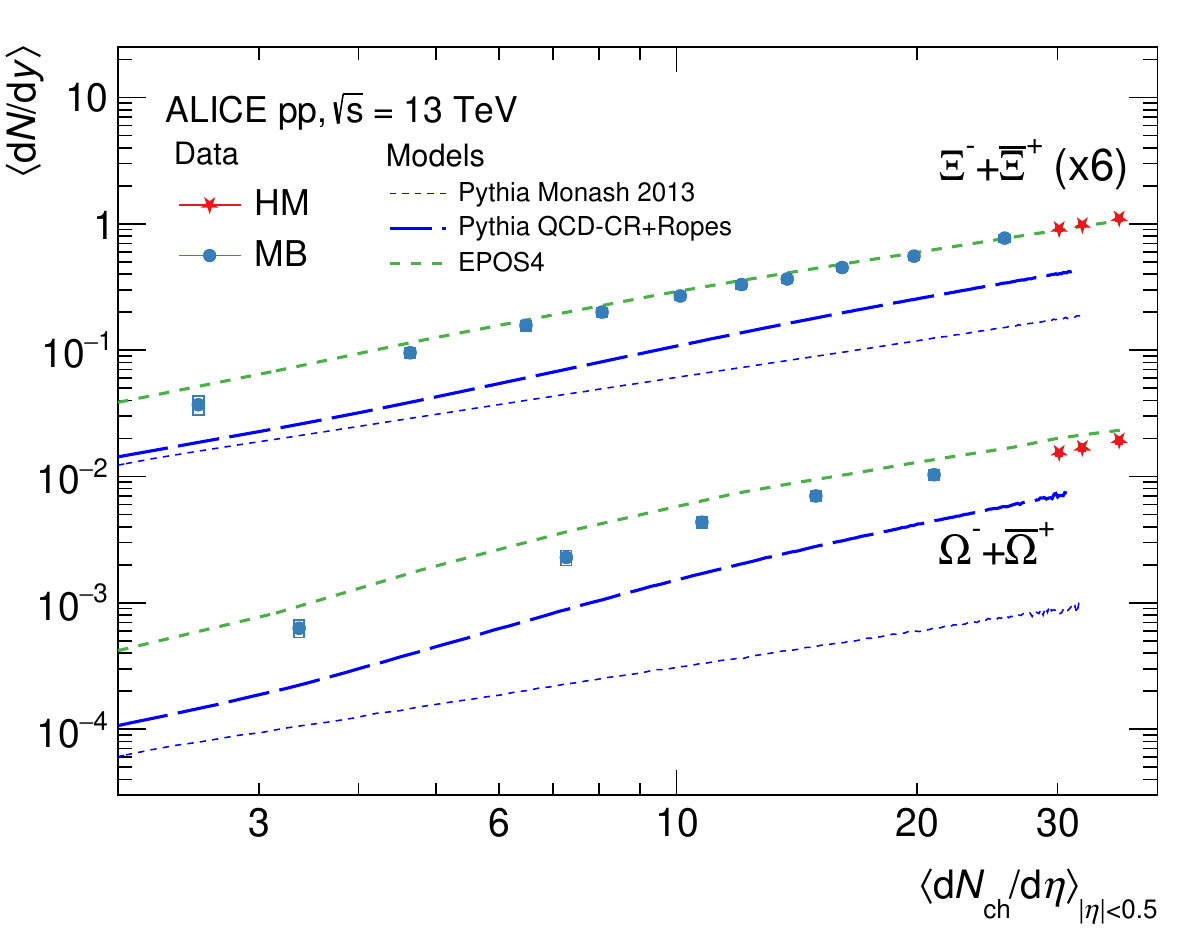}
  \caption{$\Xi$ and $\Omega$ $\pt$-integrated yields per event as a function of charged-particle multiplicity measured at midrapidity in pp collision~\cite{ALICE:2019avo}, compared to models~\cite{Sjostrand:2014zea, Bierlich:2014xba, EPOS4_PhysRevC.108.034904, EPOS4_PhysRevC.108.064903, EPOS4_PhysRevC.109.014910}. The red markers show the results in the three HM classes presented in this article. Error bars and boxes represent statistical and systematic uncertainties, respectively.
  }
  \label{int_yield_model}
\end{figure}

The yields measured in pp collisions at $\sqrt{s}= 13$~TeV are compared to the model predictions in Fig.~\ref{int_yield_model}. Both \textsc{Pythia}8.2 with the Monash 2013 tune and with Ropes underestimate the yields over the whole multiplicity interval covered by the measurement.
The underestimation of \textsc{Pythia}8.2 with the Monash tune worsens with increasing multiplicity, ranging from 60\% (80\%) at low multiplicity (\dndetashort~$\sim$~3) up to 80\% (95\%) at high multiplicity (\dndetashort $\sim$ 30) for $\Xis$ ($\Omega$). \textsc{Pythia}8.2 with Ropes underestimates the $\Xi$ yields by about 50 to 60\%, and the \Oms yields by about 55 to 65\%, rather independently of the event multiplicity, thus providing a better description of the data with respect to the Monash tune. Finally, EPOS4 provides a good description of the integrated yields: the \Xis yields are compatible with the measured ones except in the smallest multiplicity class, where they are slightly overestimated. The increase with multiplicity of the \Oms yields is well reproduced by EPOS4, which, however, overestimates them by approximately 20\% over the whole multiplicity range.

\begin{figure}[htb!]
  \centering
  \includegraphics[width=1\textwidth]{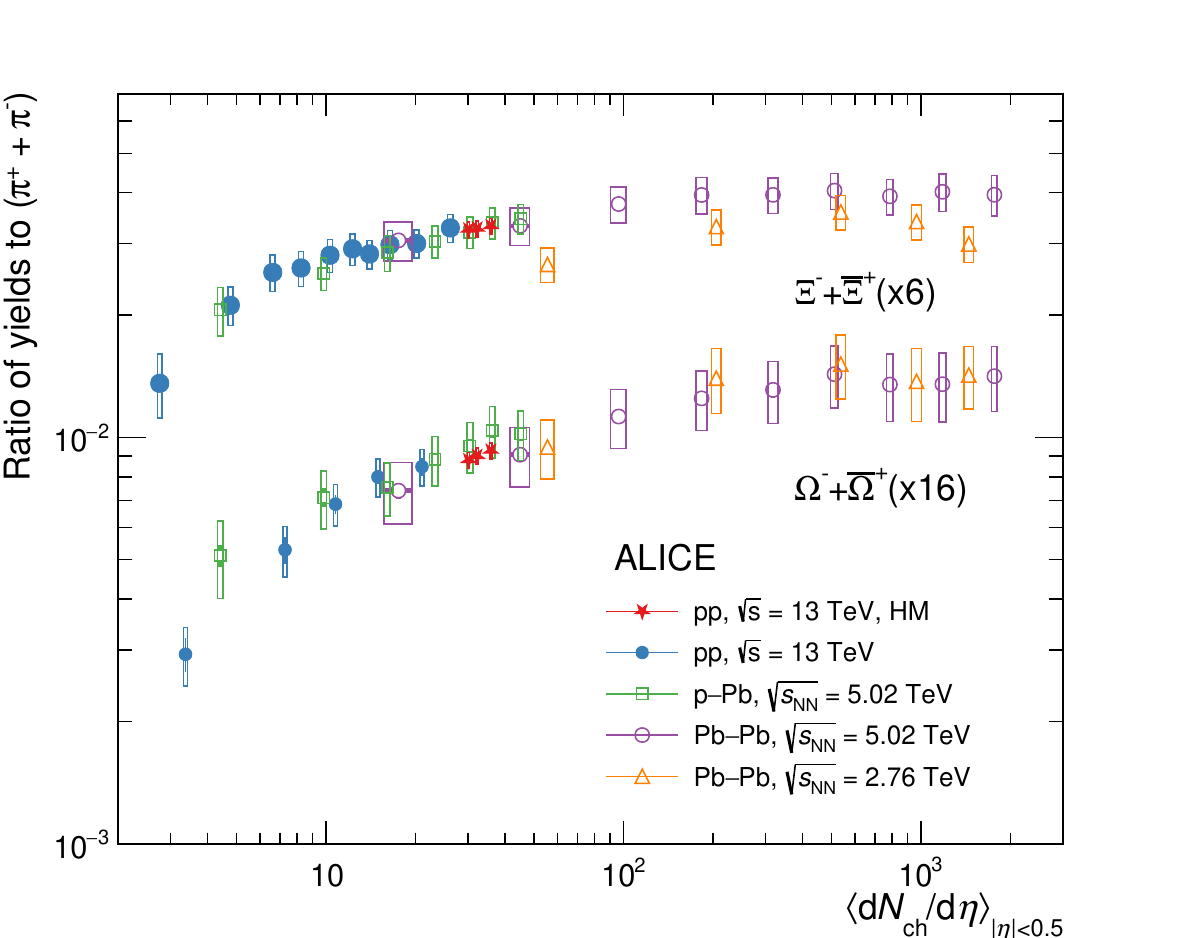}
  \caption{$\Xi/\pi$ and $\Omega/\pi$ ratios as a function of charged-particle multiplicity measured at midrapidity in pp, p--Pb and Pb--Pb collisions, as indicated in the legend~\cite{ALICE:2019avo, pPb5TeV, PbPb276TeV, PbPb5TeV}. The red markers show the results in the three HM classes presented in this article.
    Error bars and boxes represent statistical and systematic uncertainties, respectively.
  }
  \label{PionRatio}
\end{figure}

Strangeness enhancement is studied by considering the hyperon to pion ratios.
The ratios of $\Xi$ and $\Omega$ yields to pion yields are shown in Fig.~\ref{PionRatio}.
The pion yields are taken from~\cite{Pion} and extrapolated to higher multiplicity using a straight line, which provides a good description of the measured values. The systematic uncertainty of the extrapolation was calculated using the upper and lower uncertainties of the measured values at lower multiplicities.
The ratios in different collision systems at nearly identical values of multiplicities agree within $1\sigma$, with $\sigma$ given by the sum in quadrature of statistical and systematic uncertainties.
This clearly indicates that strangeness enhancement with multiplicity does not depend on collision energy or size, and is strongly correlated with the final-state multiplicity. It is worth observing that the results in HM pp collisions do not rule out the possibility of saturation of the hyperon to pion ratios with increasing multiplicity in small systems.

\begin{figure}[htb!]
  \centering
  \includegraphics[width=1\textwidth]{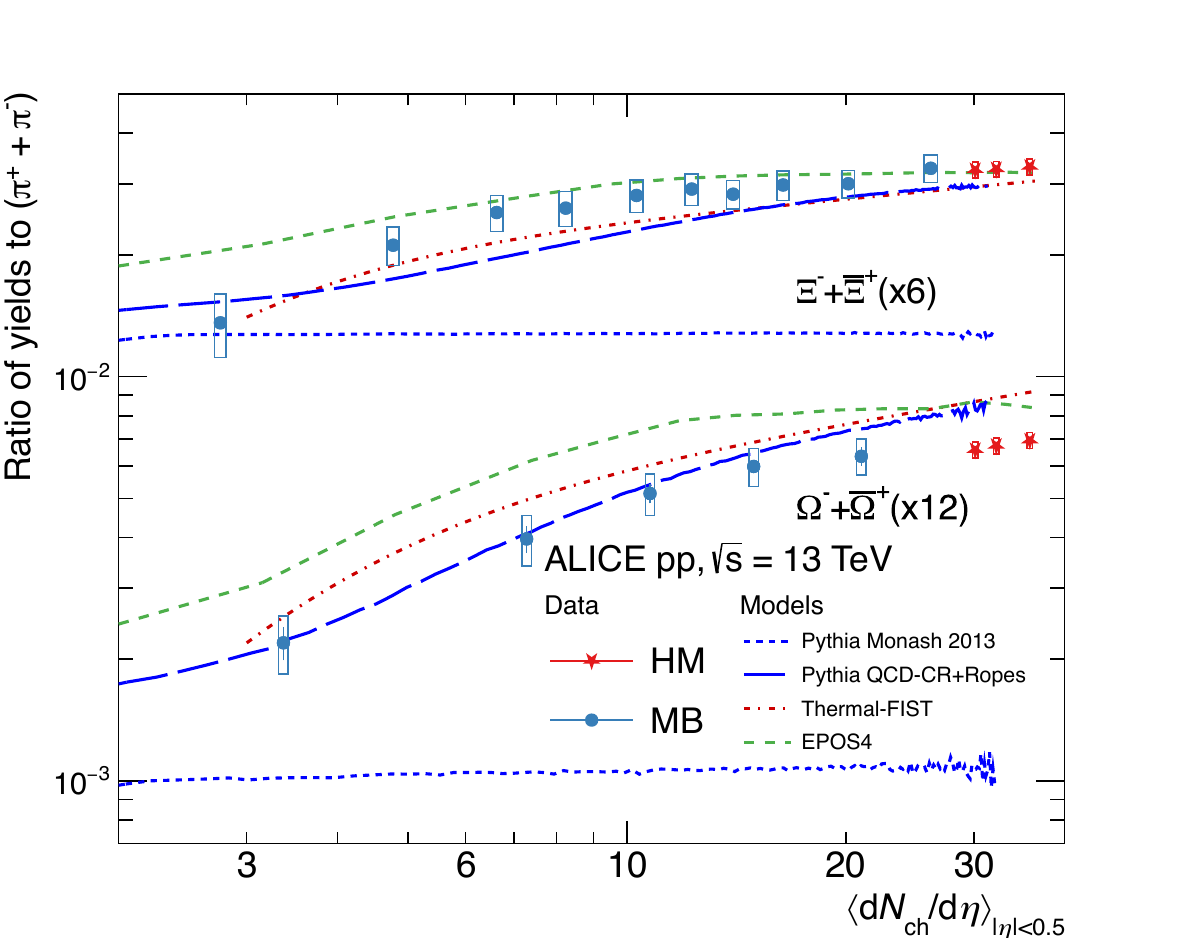}
  \caption{$\Xi/\pi$ and $\Omega/\pi$ ratios as a function of charged-particle multiplicity measured at midrapidity in pp collisions~\cite{ALICE:2019avo}, compared to models~\cite{Sjostrand:2014zea, Bierlich:2014xba, EPOS4_PhysRevC.108.034904, EPOS4_PhysRevC.108.064903, EPOS4_PhysRevC.109.014910, Vovchenko:2019kes, Vovchenko:2019pjl}. The red markers show the results in the three HM classes presented in this article. The data points are shown with markers, and the model predictions are drawn with lines of different styles, as indicated in the legend. Error bars and boxes represent statistical and systematic uncertainties, respectively.}
  \label{RatioModels}
\end{figure}

The hyperon to pion ratios in pp collisions at $\sqrt{s}=13$~TeV are compared to model predictions in Fig.~\ref{RatioModels}. In addition to the models presented above, the predictions of the canonical statistical model obtained with the Thermal FIST package are shown~\cite{Vovchenko:2019kes, Vovchenko:2019pjl}.
As already observed in Ref.~\cite{2017}, \textsc{Pythia}8.2 with the Monash tune largely underestimates the ratios and does not reproduce the increasing trend with multiplicity. On the contrary, \textsc{Pythia}8.2 with Ropes, EPOS4, and the thermal model predict strangeness enhancement with multiplicity and show a steeper increase for $\Omega$ than for $\Xi$.
\textsc{Pythia}8.2 with Ropes systematically underestimates both the $\Xi/\pi$ and $\Omega/\pi$ ratios by about 2$\sigma$. Also, the thermal model underestimates the $\Xi/\pi$ by approximately 2$\sigma$, whereas it overestimates the $\Omega/\pi$ all over the multiplicity range, and it is not compatible with the high multiplicity values (\dndetashort $\sim$ 30), which are overestimated by 30\%. The predictions of EPOS4 are compatible with the $\Xi/\pi$ ($\Omega/\pi$) ratio within 1.2$\sigma$ in the range $5<$\dndetashort$<30$ ($7<$\dndetashort$<20$), whereas the $\Omega/\pi$ ratios are overestimated by about 3 to 4$\sigma$ at the lowest and highest multiplicity values. Despite these deviations, the overall trends observed in EPOS4, \textsc{Pythia}8.2 with Ropes, and the Thermal FIST model align well with the data, suggesting that a reasonable description of the strangeness enhancement across multiplicities is achieved both with the statistical model and with QCD-based models that include collective effects and partonic interactions. The improvement of PYTHIA with ropes from that without ropes underscores the significant role of string interactions in strange baryon production within the model.

\section{Conclusions}
\label{sec:conclusions}

The production of $\Xi$ and $\Omega$ hyperons was measured in pp collisions characterised by values of multiplicity comparable to those reached in peripheral Pb--Pb collisions, with the purpose of investigating the strangeness enhancement phenomenon in small and large collision systems with similar final-state multiplicity.

The transverse momentum spectra of $\Xi$ and $\Omega$ were measured in three high-multiplicity classes: at higher multiplicity, a hardening of the spectra is observed, following the trend suggested by previous measurements in lower multiplicity pp collisions~\cite{7TeV, ALICE:2019avo}. The comparison with model predictions shows that \textsc{Pythia}8.2 with the Monash 2013 tune significantly underestimates spectra yields, while \textsc{Pythia}8.2 with Ropes improves the description at low \pt but predicts softer spectra at intermediate \pt. EPOS4 provides the best description of the $\Xi$ and $\Omega$ spectra shape, though it slightly underestimates the yields. Overall, all models tend to underestimate the high-\pt yields for both particles.

The \meanpt of $\Xi$ in the highest multiplicity pp collisions is observed to be larger than the one measured in p--Pb collisions with similar multiplicity, whereas previous measurements show comparable results between the two systems for multiplicities $\dndetashort < 20$. 
The different behaviour between the two systems for $\dndetashort > 20$ might be related to the different initial states and energy densities in pp and p--Pb collisions with the same multiplicity~\cite{ALICE:2013rdo}. Predictions from \textsc{Pythia}8.2 with either the Monash 2013 tune or the Ropes mechanism capture the increase of \meanpt with the multiplicity but underestimate the values. EPOS4 reproduces the \meanpt trends, with a good agreement at low multiplicity for $\Xi$ and a slight underestimation at the highest multiplicities. On the contrary, the \meanpt of $\Omega$ is well described at high multiplicity but slightly overestimated at intermediate multiplicity.

The \pt-integrated yields exhibit a rising trend with increasing multiplicity, consistent with expectations based on previous measurements in pp collisions at lower multiplicities.
Comparison with other systems, such as p--Pb and Pb--Pb collisions at different collision energies, shows agreement of the yields within uncertainties at similar multiplicity values.
This observation suggests that hyperon production is independent of the initial system size and is strongly correlated with the final-state multiplicity.
Both \textsc{Pythia}8.2 with the Monash 2013 tune and with Ropes underestimate the yields over the whole multiplicity interval covered by the measurement. The underestimation is smaller for \textsc{Pythia}8.2 with Ropes, which is able to reproduce the relative increase with multiplicity fairly well. EPOS4 provides a generally accurate description of the integrated yields. The $\Xi$ yields are well reproduced across multiplicity classes, with only a slight overestimation in the lowest multiplicity class. The multiplicity-dependent increase of the $\Omega$ yields is correctly captured, although EPOS4 systematically overestimates them by about 20\% across the entire multiplicity range.

The ratios of hyperon to charged-pion yields provide a probe of strangeness enhancement with multiplicity.
The relative hyperon production shows a quickly growing trend at lower multiplicities, with saturation occurring in central Pb--Pb collisions.
The measurements in high-multiplicity pp collisions presented here follow the increasing trend and are consistent with the p--Pb and Pb--Pb ones at the same multiplicity, confirming the strong correlation between strangeness enhancement and final-state multiplicity, regardless of initial system size or energy.
The ratios in high-multiplicity pp collisions show a hint of increase with multiplicity, but systematic uncertainties prevent ruling out the possible reach of a saturation value.
While \textsc{Pythia}8.2 with the Monash tune largely underestimates the ratios and does not reproduce the increasing trend with multiplicity, \textsc{Pythia}8.2 with Ropes, EPOS4, and the thermal model predict fairly well the strangeness enhancement with multiplicity, which shows a steeper increase for $\Omega$ than for $\Xi$.

Further investigation of the origin of strangeness enhancement in high-multiplicity pp collisions will be possible thanks to the record sample of pp collisions collected during Run~3, which is expected to be approximately twenty times larger than the sample used in this analysis.
The multi-strange hadron yields in Run 3 data can be measured up to significantly larger values of multiplicity in pp collisions, potentially revealing whether the $\Xi/\pi$ and $\Omega/\pi$ ratios in the pp system reach, or possibly exceed, the thermal limit observed in central Pb--Pb collisions.


\newenvironment{acknowledgement}{\relax}{\relax}
\begin{acknowledgement}
\section*{Acknowledgements}
\input{fa_2026-02-19_Opt_C.tex}
\end{acknowledgement}

\bibliographystyle{utphys}   
\bibliography{bibliography}

\newpage
\appendix

%
%

\section{The ALICE Collaboration}
\label{app:collab}
\input{Alice_Authorlist_2026-02-19_Opt_C.tex}  
\end{document}

%% file: commands.tex
%

\newcommand{\pp}           {pp\xspace}
\newcommand{\ppbar}        {\mbox{$\mathrm {p\overline{p}}$}\xspace}
\newcommand{\XeXe}         {\mbox{Xe--Xe}\xspace}
\newcommand{\PbPb}         {\mbox{Pb--Pb}\xspace}
\newcommand{\pA}           {\mbox{pA}\xspace}
\newcommand{\pPb}          {\mbox{p--Pb}\xspace}
\newcommand{\AuAu}         {\mbox{Au--Au}\xspace}
\newcommand{\dAu}          {\mbox{d--Au}\xspace}

\newcommand{\s}            {\ensuremath{\sqrt{s}}\xspace}
\newcommand{\snn}          {\ensuremath{\sqrt{s_{\mathrm{NN}}}}\xspace}
\newcommand{\pt}           {\ensuremath{p_{\rm T}}\xspace}
\newcommand{\meanpt}       {$\langle p_{\mathrm{T}}\rangle$\xspace}
\newcommand{\ycms}         {\ensuremath{y_{\rm CMS}}\xspace}
\newcommand{\ylab}         {\ensuremath{y_{\rm lab}}\xspace}
\newcommand{\etarange}[1]  {\mbox{$\left | \eta \right |~<~#1$}}
\newcommand{\yrange}[1]    {\mbox{$\left | y \right |~<~#1$}}
\newcommand{\dndy}         {\ensuremath{\mathrm{d}N_\mathrm{ch}/\mathrm{d}y}\xspace}
\newcommand{\dndeta}       {\ensuremath{\mathrm{d}N_\mathrm{ch}/\mathrm{d}\eta}\xspace}
\newcommand{\avdndeta}     {\ensuremath{\langle\dndeta\rangle}\xspace}
\newcommand{\dNdy}         {\ensuremath{\mathrm{d}N_\mathrm{ch}/\mathrm{d}y}\xspace}
\newcommand{\Npart}        {\ensuremath{N_\mathrm{part}}\xspace}
\newcommand{\Ncoll}        {\ensuremath{N_\mathrm{coll}}\xspace}
\newcommand{\dEdx}         {\ensuremath{\textrm{d}E/\textrm{d}x}\xspace}
\newcommand{\RpPb}         {\ensuremath{R_{\rm pPb}}\xspace}

\newcommand{\nineH}        {$\sqrt{s}~=~0.9$~Te\kern-.1emV\xspace}
\newcommand{\seven}        {$\sqrt{s}~=~7$~Te\kern-.1emV\xspace}
\newcommand{\twoH}         {$\sqrt{s}~=~0.2$~Te\kern-.1emV\xspace}
\newcommand{\twosevensix}  {$\sqrt{s}~=~2.76$~Te\kern-.1emV\xspace}
\newcommand{\five}         {$\sqrt{s}~=~5.02$~Te\kern-.1emV\xspace}
\newcommand{\twosevensixnn}{$\sqrt{s_{\mathrm{NN}}}~=~2.76$~Te\kern-.1emV\xspace}
\newcommand{\fivenn}       {$\sqrt{s_{\mathrm{NN}}}~=~5.02$~Te\kern-.1emV\xspace}
\newcommand{\LT}           {L{\'e}vy-Tsallis\xspace}
\newcommand{\GeVc}         {Ge\kern-.1emV/$c$\xspace}
\newcommand{\MeVc}         {Me\kern-.1emV/$c$\xspace}
\newcommand{\TeV}          {Te\kern-.1emV\xspace}
\newcommand{\GeV}          {Ge\kern-.1emV\xspace}
\newcommand{\MeV}          {Me\kern-.1emV\xspace}
\newcommand{\GeVmass}      {Ge\kern-.2emV/$c^2$\xspace}
\newcommand{\MeVmass}      {Me\kern-.2emV/$c^2$\xspace}
\newcommand{\lumi}         {\ensuremath{\mathcal{L}}\xspace}

\newcommand{\ITS}          {\rm{ITS}\xspace}
\newcommand{\TOF}          {\rm{TOF}\xspace}
\newcommand{\ZDC}          {\rm{ZDC}\xspace}
\newcommand{\ZDCs}         {\rm{ZDCs}\xspace}
\newcommand{\ZNA}          {\rm{ZNA}\xspace}
\newcommand{\ZNC}          {\rm{ZNC}\xspace}
\newcommand{\SPD}          {\rm{SPD}\xspace}
\newcommand{\SDD}          {\rm{SDD}\xspace}
\newcommand{\SSD}          {\rm{SSD}\xspace}
\newcommand{\TPC}          {\rm{TPC}\xspace}
\newcommand{\TRD}          {\rm{TRD}\xspace}
\newcommand{\VZERO}        {\rm{V0}\xspace}
\newcommand{\VZEROA}       {\rm{V0A}\xspace}
\newcommand{\VZEROC}       {\rm{V0C}\xspace}
\newcommand{\Vdecay} 	   {\ensuremath{V^{0}}\xspace}

\newcommand{\ee}           {\ensuremath{e^{+}e^{-}}} 
\newcommand{\pip}          {\ensuremath{\pi^{+}}\xspace}
\newcommand{\pim}          {\ensuremath{\pi^{-}}\xspace}
\newcommand{\kap}          {\ensuremath{\rm{K}^{+}}\xspace}
\newcommand{\kam}          {\ensuremath{\rm{K}^{-}}\xspace}
\newcommand{\pbar}         {\ensuremath{\rm\overline{p}}\xspace}
\newcommand{\kzero}        {\ensuremath{{\rm K}^{0}_{\rm{S}}}\xspace}
\newcommand{\lmb}          {\ensuremath{\Lambda}\xspace}
\newcommand{\almb}         {\ensuremath{\overline{\Lambda}}\xspace}
\newcommand{\Om}           {\ensuremath{\Omega^-}\xspace}
\newcommand{\Mo}           {\ensuremath{\overline{\Omega}^+}\xspace}
\newcommand{\X}            {\ensuremath{\Xi^-}\xspace}
\newcommand{\Ix}           {\ensuremath{\overline{\Xi}^+}\xspace}
\newcommand{\Xis}         {\ensuremath{\Xi}\xspace}
\newcommand{\Oms}         {\ensuremath{\Omega}\xspace}
\newcommand{\degree}       {\ensuremath{^{\rm o}}\xspace}

\newcommand{\xim}{\ensuremath{{\rm \Xi}^{-}}\xspace}
\newcommand{\xip}{\ensuremath{{\rm \Bar{\Xi}}^{+}}\xspace}

\newcommand{\omm}{\ensuremath{{\rm \Omega}^{-}}\xspace}
\newcommand{\omp}{\ensuremath{{\rm \Bar{\Omega}}^{+}}\xspace}
\newcommand{\dndetashort} {\ensuremath{\langle\dndeta\rangle}\xspace}

%% file: fa_2026-02-19_Opt_C.tex

The ALICE Collaboration would like to thank all its engineers and technicians for their invaluable contributions to the construction of the experiment and the CERN accelerator teams for the outstanding performance of the LHC complex.
The ALICE Collaboration gratefully acknowledges the resources and support provided by all Grid centres and the Worldwide LHC Computing Grid (WLCG) collaboration.
The ALICE Collaboration acknowledges the following funding agencies for their support in building and running the ALICE detector:
A. I. Alikhanyan National Science Laboratory (Yerevan Physics Institute) Foundation (ANSL), State Committee of Science and World Federation of Scientists (WFS), Armenia;
Austrian Academy of Sciences, Austrian Science Fund (FWF): [M 2467-N36] and Nationalstiftung f\"{u}r Forschung, Technologie und Entwicklung, Austria;
Ministry of Communications and High Technologies, National Nuclear Research Center, Azerbaijan;
Rede Nacional de Física de Altas Energias (Renafae), Financiadora de Estudos e Projetos (Finep), Funda\c{c}\~{a}o de Amparo \`{a} Pesquisa do Estado de S\~{a}o Paulo (FAPESP) and The Sao Paulo Research Foundation  (FAPESP), Brazil;
Bulgarian Ministry of Education and Science, within the National Roadmap for Research Infrastructures 2020-2027 (object CERN), Bulgaria;
Ministry of Education of China (MOEC) , Ministry of Science \& Technology of China (MSTC) and National Natural Science Foundation of China (NSFC), China;
Ministry of Science and Education and Croatian Science Foundation, Croatia;
Centro de Aplicaciones Tecnol\'{o}gicas y Desarrollo Nuclear (CEADEN), Cubaenerg\'{\i}a, Cuba;
Ministry of Education, Youth and Sports of the Czech Republic, Czech Republic;
The Danish Council for Independent Research | Natural Sciences, the VILLUM FONDEN and Danish National Research Foundation (DNRF), Denmark;
Helsinki Institute of Physics (HIP), Finland;
Commissariat \`{a} l'Energie Atomique (CEA) and Institut National de Physique Nucl\'{e}aire et de Physique des Particules (IN2P3) and Centre National de la Recherche Scientifique (CNRS), France;
Bundesministerium f\"{u}r Forschung, Technologie und Raumfahrt (BMFTR) and GSI Helmholtzzentrum f\"{u}r Schwerionenforschung GmbH, Germany;
National Research, Development and Innovation Office, Hungary;
Department of Atomic Energy Government of India (DAE), Department of Science and Technology, Government of India (DST), University Grants Commission, Government of India (UGC) and Council of Scientific and Industrial Research (CSIR), India;
National Research and Innovation Agency - BRIN, Indonesia;
Istituto Nazionale di Fisica Nucleare (INFN), Italy;
Japanese Ministry of Education, Culture, Sports, Science and Technology (MEXT) and Japan Society for the Promotion of Science (JSPS) KAKENHI, Japan;
Consejo Nacional de Ciencia (CONACYT) y Tecnolog\'{i}a, through Fondo de Cooperaci\'{o}n Internacional en Ciencia y Tecnolog\'{i}a (FONCICYT) and Direcci\'{o}n General de Asuntos del Personal Academico (DGAPA), Mexico;
Nederlandse Organisatie voor Wetenschappelijk Onderzoek (NWO), Netherlands;
The Research Council of Norway, Norway;
Pontificia Universidad Cat\'{o}lica del Per\'{u}, Peru;
Ministry of Science and Higher Education, National Science Centre and WUT ID-UB, Poland;
Korea Institute of Science and Technology Information and National Research Foundation of Korea (NRF), Republic of Korea;
Ministry of Education and Scientific Research, Institute of Atomic Physics, Ministry of Research and Innovation and Institute of Atomic Physics and Universitatea Nationala de Stiinta si Tehnologie Politehnica Bucuresti, Romania;
Ministerstvo skolstva, vyskumu, vyvoja a mladeze SR, Slovakia;
National Research Foundation of South Africa, South Africa;
Swedish Research Council (VR) and Knut \& Alice Wallenberg Foundation (KAW), Sweden;
European Organization for Nuclear Research, Switzerland;
Suranaree University of Technology (SUT), National Science and Technology Development Agency (NSTDA) and National Science, Research and Innovation Fund (NSRF via PMU-B B05F650021), Thailand;
Turkish Energy, Nuclear and Mineral Research Agency (TENMAK), Turkey;
National Academy of  Sciences of Ukraine, Ukraine;
Science and Technology Facilities Council (STFC), United Kingdom;
National Science Foundation of the United States of America (NSF) and United States Department of Energy, Office of Nuclear Physics (DOE NP), United States of America.
In addition, individual groups or members have received support from:
FORTE project, reg.\ no.\ CZ.02.01.01/00/22\_008/0004632, Czech Republic, co-funded by the European Union, Czech Republic;
European Research Council (grant no. 950692), European Union;
Deutsche Forschungs Gemeinschaft (DFG, German Research Foundation) ``Neutrinos and Dark Matter in Astro- and Particle Physics'' (grant no. SFB 1258), Germany;
FAIR - Future Artificial Intelligence Research, funded by the NextGenerationEU program (Italy).

%% file: Alice_Authorlist_2026-02-19_Opt_C.tex
\begin{flushleft} 
\small

D.A.H.~Abdallah\,\orcidlink{0000-0003-4768-2718}\,$^{\rm 134}$, 
I.J.~Abualrob\,\orcidlink{0009-0005-3519-5631}\,$^{\rm 112}$, 
S.~Acharya\,\orcidlink{0000-0002-9213-5329}\,$^{\rm 49}$, 
K.~Agarwal\,\orcidlink{0000-0001-5781-3393}\,$^{\rm II,}$$^{\rm 23}$, 
G.~Aglieri Rinella\,\orcidlink{0000-0002-9611-3696}\,$^{\rm 32}$, 
L.~Aglietta\,\orcidlink{0009-0003-0763-6802}\,$^{\rm 24}$, 
N.~Agrawal\,\orcidlink{0000-0003-0348-9836}\,$^{\rm 25}$, 
Z.~Ahammed\,\orcidlink{0000-0001-5241-7412}\,$^{\rm 132}$, 
S.~Ahmad\,\orcidlink{0000-0003-0497-5705}\,$^{\rm 15}$, 
I.~Ahuja\,\orcidlink{0000-0002-4417-1392}\,$^{\rm 36}$
% ...

\section*{Affiliation Notes}

$^{\rm I}$ Deceased\\
$^{\rm II}$ Also at: INFN Trieste\\
$^{\rm III}$ Also at: Fondazione Bruno Kessler (FBK), Trento, Italy\\
$^{\rm IV}$ Also at: Max-Planck-Institut fur Physik, Munich, Germany\\
$^{\rm V}$ Also at: Czech Technical University in Prague (CZ)\\
$^{\rm VI}$ Also at: Instituto de Fisica da Universidade de Sao Paulo\\
$^{\rm VII}$ Also at: Dipartimento DET del Politecnico di Torino, Turin, Italy\\
$^{\rm VIII}$ Also at: Department of Applied Physics, Aligarh Muslim University, Aligarh, India\\
$^{\rm IX}$ Also at: Institute of Theoretical Physics, University of Wroclaw, Poland\\
$^{\rm X}$ Also at: Facultad de Ciencias, Universidad Nacional Aut\'{o}noma de M\'{e}xico, Mexico City, Mexico\\

\section*{Collaboration Institutes}

$^{1}$ A.I. Alikhanyan National Science Laboratory (Yerevan Physics Institute) Foundation, Yerevan, Armenia\\
$^{2}$ AGH University of Krakow, Cracow, Poland\\
$^{3}$ Bogolyubov Institute for Theoretical Physics, National Academy of Sciences of Ukraine, Kyiv, Ukraine\\
$^{4}$ Bose Institute, Department of Physics  and Centre for Astroparticle Physics and Space Science (CAPSS), Kolkata, India\\
$^{5}$ California Polytechnic State University, San Luis Obispo, California, United States\\
$^{6}$ Central China Normal University, Wuhan, China\\
$^{7}$ Centro de Aplicaciones Tecnol\'{o}gicas y Desarrollo Nuclear (CEADEN), Havana, Cuba\\
$^{8}$ Centro de Investigaci\'{o}n y de Estudios Avanzados (CINVESTAV), Mexico City and M\'{e}rida, Mexico\\
$^{9}$ Chicago State University, Chicago, Illinois, United States\\
$^{10}$ China Nuclear Data Center, China Institute of Atomic Energy, Beijing, China\\
$^{11}$ China University of Geosciences, Wuhan, China\\
$^{12}$ Chungbuk National University, Cheongju, Republic of Korea\\
$^{13}$ Comenius University Bratislava, Faculty of Mathematics, Physics and Informatics, Bratislava, Slovak Republic\\
$^{14}$ Creighton University, Omaha, Nebraska, United States\\
$^{15}$ Department of Physics, Aligarh Muslim University, Aligarh, India\\
$^{16}$ Department of Physics, Pusan National University, Pusan, Republic of Korea\\
$^{17}$ Department of Physics, Sejong University, Seoul, Republic of Korea\\
$^{18}$ Department of Physics, University of California, Berkeley, California, United States\\
$^{19}$ Department of Physics, University of Oslo, Oslo, Norway\\
$^{20}$ Department of Physics and Technology, University of Bergen, Bergen, Norway\\
$^{21}$ Dipartimento di Fisica, Universit\`{a} di Pavia, Pavia, Italy\\
$^{22}$ Dipartimento di Fisica dell'Universit\`{a} and Sezione INFN, Cagliari, Italy\\
$^{23}$ Dipartimento di Fisica dell'Universit\`{a} and Sezione INFN, Trieste, Italy\\
$^{24}$ Dipartimento di Fisica dell'Universit\`{a} and Sezione INFN, Turin, Italy\\
$^{25}$ Dipartimento di Fisica e Astronomia dell'Universit\`{a} and Sezione INFN, Bologna, Italy\\
$^{26}$ Dipartimento di Fisica e Astronomia dell'Universit\`{a} and Sezione INFN, Catania, Italy\\
$^{27}$ Dipartimento di Fisica e Astronomia dell'Universit\`{a} and Sezione INFN, Padova, Italy\\
$^{28}$ Dipartimento di Fisica `E.R.~Caianiello' dell'Universit\`{a} and Gruppo Collegato INFN, Salerno, Italy\\
$^{29}$ Dipartimento DISAT del Politecnico and Sezione INFN, Turin, Italy\\
$^{30}$ Dipartimento di Scienze MIFT, Universit\`{a} di Messina, Messina, Italy\\
$^{31}$ Dipartimento Interateneo di Fisica `M.~Merlin' and Sezione INFN, Bari, Italy\\
$^{32}$ European Organization for Nuclear Research (CERN), Geneva, Switzerland\\
$^{33}$ Faculty of Electrical Engineering, Mechanical Engineering and Naval Architecture, University of Split, Split, Croatia\\
$^{34}$ Faculty of Nuclear Sciences and Physical Engineering, Czech Technical University in Prague, Prague, Czech Republic\\
$^{35}$ Faculty of Physics, Sofia University, Sofia, Bulgaria\\
$^{36}$ Faculty of Science, P.J.~\v{S}af\'{a}rik University, Ko\v{s}ice, Slovak Republic\\
$^{37}$ Faculty of Technology, Environmental and Social Sciences, Bergen, Norway\\
$^{38}$ Frankfurt Institute for Advanced Studies, Johann Wolfgang Goethe-Universit\"{a}t Frankfurt, Frankfurt, Germany\\
$^{39}$ Fudan University, Shanghai, China\\
$^{40}$ Gauhati University, Department of Physics, Guwahati, India\\
$^{41}$ Helmholtz-Institut f\"{u}r Strahlen- und Kernphysik, Rheinische Friedrich-Wilhelms-Universit\"{a}t Bonn, Bonn, Germany\\
$^{42}$ Helsinki Institute of Physics (HIP), Helsinki, Finland\\
$^{43}$ High Energy Physics Group,  Universidad Aut\'{o}noma de Puebla, Puebla, Mexico\\
$^{44}$ Horia Hulubei National Institute of Physics and Nuclear Engineering, Bucharest, Romania\\
$^{45}$ HUN-REN Wigner Research Centre for Physics, Budapest, Hungary\\
$^{46}$ Indian Institute of Technology Bombay (IIT), Mumbai, India\\
$^{47}$ Indian Institute of Technology Indore, Indore, India\\
$^{48}$ INFN, Laboratori Nazionali di Frascati, Frascati, Italy\\
$^{49}$ INFN, Sezione di Bari, Bari, Italy\\
$^{50}$ INFN, Sezione di Bologna, Bologna, Italy\\
$^{51}$ INFN, Sezione di Cagliari, Cagliari, Italy\\
$^{52}$ INFN, Sezione di Catania, Catania, Italy\\
$^{53}$ INFN, Sezione di Padova, Padova, Italy\\
$^{54}$ INFN, Sezione di Pavia, Pavia, Italy\\
$^{55}$ INFN, Sezione di Torino, Turin, Italy\\
$^{56}$ INFN, Sezione di Trieste, Trieste, Italy\\
$^{57}$ Inha University, Incheon, Republic of Korea\\
$^{58}$ Institute for Gravitational and Subatomic Physics (GRASP), Utrecht University/Nikhef, Utrecht, Netherlands\\
$^{59}$ Institute of Experimental Physics, Slovak Academy of Sciences, Ko\v{s}ice, Slovak Republic\\
$^{60}$ Institute of Physics, Homi Bhabha National Institute, Bhubaneswar, India\\
$^{61}$ Institute of Physics of the Czech Academy of Sciences, Prague, Czech Republic\\
$^{62}$ Institute of Space Science (ISS), Bucharest, Romania\\
$^{63}$ Institut f\"{u}r Kernphysik, Johann Wolfgang Goethe-Universit\"{a}t Frankfurt, Frankfurt, Germany\\
$^{64}$ Instituto de Ciencias Nucleares, Universidad Nacional Aut\'{o}noma de M\'{e}xico, Mexico City, Mexico\\
$^{65}$ Instituto de F\'{i}sica, Universidade Federal do Rio Grande do Sul (UFRGS), Porto Alegre, Brazil\\
$^{66}$ Instituto de F\'{\i}sica, Universidad Nacional Aut\'{o}noma de M\'{e}xico, Mexico City, Mexico\\
$^{67}$ iThemba LABS, National Research Foundation, Somerset West, South Africa\\
$^{68}$ Jeonbuk National University, Jeonju, Republic of Korea\\
$^{69}$ Korea Institute of Science and Technology Information, Daejeon, Republic of Korea\\
$^{70}$ Laboratoire de Physique Subatomique et de Cosmologie, Universit\'{e} Grenoble-Alpes, CNRS-IN2P3, Grenoble, France\\
$^{71}$ Lawrence Berkeley National Laboratory, Berkeley, California, United States\\
$^{72}$ Lund University Department of Physics, Division of Particle Physics, Lund, Sweden\\
$^{73}$ Marietta Blau Institute, Vienna, Austria\\
$^{74}$ Nagasaki Institute of Applied Science, Nagasaki, Japan\\
$^{75}$ Nara Women{'}s University (NWU), Nara, Japan\\
$^{76}$ National Centre for Nuclear Research, Warsaw, Poland\\
$^{77}$ National Institute of Science Education and Research, Homi Bhabha National Institute, Jatni, India\\
$^{78}$ National Nuclear Research Center, Baku, Azerbaijan\\
$^{79}$ National Research and Innovation Agency - BRIN, Jakarta, Indonesia\\
$^{80}$ Niels Bohr Institute, University of Copenhagen, Copenhagen, Denmark\\
$^{81}$ Nikhef, National institute for subatomic physics, Amsterdam, Netherlands\\
$^{82}$ Nuclear Physics Group, STFC Daresbury Laboratory, Daresbury, United Kingdom\\
$^{83}$ Nuclear Physics Institute of the Czech Academy of Sciences, Husinec-\v{R}e\v{z}, Czech Republic\\
$^{84}$ Oak Ridge National Laboratory, Oak Ridge, Tennessee, United States\\
$^{85}$ Ohio State University, Columbus, Ohio, United States\\
$^{86}$ Physics department, Faculty of science, University of Zagreb, Zagreb, Croatia\\
$^{87}$ Physics Department, Panjab University, Chandigarh, India\\
$^{88}$ Physics Department, University of Jammu, Jammu, India\\
$^{89}$ Physics Program and International Institute for Sustainability with Knotted Chiral Meta Matter (WPI-SKCM$^{2}$), Hiroshima University, Hiroshima, Japan\\
$^{90}$ Physikalisches Institut, Eberhard-Karls-Universit\"{a}t T\"{u}bingen, T\"{u}bingen, Germany\\
$^{91}$ Physikalisches Institut, Ruprecht-Karls-Universit\"{a}t Heidelberg, Heidelberg, Germany\\
$^{92}$ Physik Department, Technische Universit\"{a}t M\"{u}nchen, Munich, Germany\\
$^{93}$ Politecnico di Bari and Sezione INFN, Bari, Italy\\
$^{94}$ Research Division and ExtreMe Matter Institute EMMI, GSI Helmholtzzentrum f\"ur Schwerionenforschung GmbH, Darmstadt, Germany\\
$^{95}$ Saga University, Saga, Japan\\
$^{96}$ Saha Institute of Nuclear Physics, Homi Bhabha National Institute, Kolkata, India\\
$^{97}$ School of Physics and Astronomy, University of Birmingham, Birmingham, United Kingdom\\
$^{98}$ Secci\'{o}n F\'{\i}sica, Departamento de Ciencias, Pontificia Universidad Cat\'{o}lica del Per\'{u}, Lima, Peru\\
$^{99}$ SUBATECH, IMT Atlantique, Nantes Universit\'{e}, CNRS-IN2P3, Nantes, France\\
$^{100}$ Sungkyunkwan University, Suwon City, Republic of Korea\\
$^{101}$ Suranaree University of Technology, Nakhon Ratchasima, Thailand\\
$^{102}$ Technical University of Ko\v{s}ice, Ko\v{s}ice, Slovak Republic\\
$^{103}$ The Henryk Niewodniczanski Institute of Nuclear Physics, Polish Academy of Sciences, Cracow, Poland\\
$^{104}$ The University of Texas at Austin, Austin, Texas, United States\\
$^{105}$ Universidad Aut\'{o}noma de Sinaloa, Culiac\'{a}n, Mexico\\
$^{106}$ Universidade de S\~{a}o Paulo (USP), S\~{a}o Paulo, Brazil\\
$^{107}$ Universidade Estadual de Campinas (UNICAMP), Campinas, Brazil\\
$^{108}$ Universidade Federal do ABC, Santo Andre, Brazil\\
$^{109}$ Universitatea Nationala de Stiinta si Tehnologie Politehnica Bucuresti, Bucharest, Romania\\
$^{110}$ University of Cape Town, Cape Town, South Africa\\
$^{111}$ University of Derby, Derby, United Kingdom\\
$^{112}$ University of Houston, Houston, Texas, United States\\
$^{113}$ University of Jyv\"{a}skyl\"{a}, Jyv\"{a}skyl\"{a}, Finland\\
$^{114}$ University of Kansas, Lawrence, Kansas, United States\\
$^{115}$ University of Liverpool, Liverpool, United Kingdom\\
$^{116}$ University of Science and Technology of China, Hefei, China\\
$^{117}$ University of Silesia in Katowice, Katowice, Poland\\
$^{118}$ University of South-Eastern Norway, Kongsberg, Norway\\
$^{119}$ University of Tennessee, Knoxville, Tennessee, United States\\
$^{120}$ University of the Witwatersrand, Johannesburg, South Africa\\
$^{121}$ University of Tokyo, Tokyo, Japan\\
$^{122}$ University of Tsukuba, Tsukuba, Japan\\
$^{123}$ Universit\"{a}t M\"{u}nster, Institut f\"{u}r Kernphysik, M\"{u}nster, Germany\\
$^{124}$ Universit\'{e} Clermont Auvergne, CNRS/IN2P3, LPC, Clermont-Ferrand, France\\
$^{125}$ Universit\'{e} de Lyon, CNRS/IN2P3, Institut de Physique des 2 Infinis de Lyon, Lyon, France\\
$^{126}$ Universit\'{e} de Strasbourg, CNRS, IPHC UMR 7178, F-67000 Strasbourg, France, Strasbourg, France\\
$^{127}$ Universit\'{e} Paris-Saclay, Centre d'Etudes de Saclay (CEA), IRFU, D\'{e}partment de Physique Nucl\'{e}aire (DPhN), Saclay, France\\
$^{128}$ Universit\'{e}  Paris-Saclay, CNRS/IN2P3, IJCLab, Orsay, France\\
$^{129}$ Universit\`{a} degli Studi di Foggia, Foggia, Italy\\
$^{130}$ Universit\`{a} del Piemonte Orientale, Vercelli, Italy\\
$^{131}$ Universit\`{a} di Brescia, Brescia, Italy\\
$^{132}$ Variable Energy Cyclotron Centre, Homi Bhabha National Institute, Kolkata, India\\
$^{133}$ Warsaw University of Technology, Warsaw, Poland\\
$^{134}$ Wayne State University, Detroit, Michigan, United States\\
$^{135}$ Yale University, New Haven, Connecticut, United States\\
$^{136}$ Yildiz Technical University, Istanbul, Turkey\\
$^{137}$ Yonsei University, Seoul, Republic of Korea\\
$^{138}$ Affiliated with an institute formerly covered by a cooperation agreement with CERN\\
$^{139}$ Affiliated with an international laboratory covered by a cooperation agreement with CERN.\\

\end{flushleft} 